\documentclass[12pt]{iopart}
\usepackage[utf8]{inputenc} 
\usepackage[T1]{fontenc}
\expandafter\let\csname equation*\endcsname\relax
\expandafter\let\csname endequation*\endcsname\relax
\usepackage{amsmath}
\usepackage{hyperref} 
\usepackage{cleveref} 
\hypersetup{colorlinks = true, 
	    linkcolor = blue, 
	    urlcolor = blue,
           citecolor = blue} 
\usepackage{url}            
\usepackage{booktabs} 
\usepackage{amsfonts}       
\usepackage{nicefrac}      
\usepackage{microtype}           
\usepackage[pdftex]{graphicx}       
\usepackage{subfigure}
\usepackage{color}
\usepackage{qcircuit}
\usepackage{float} 
\usepackage{amsthm}

\graphicspath{{media/}}

\usepackage{iopams} 
\begin{document}

\title[Entanglement distribution based on quantum walk in arbitrary quantum networks ]{Entanglement distribution based on quantum walk in arbitrary quantum networks }

\author{Tianen Chen$^{1,2}$, Yun Shang$^{1,3,\dag}$, Chitong Chen$^{4,5}$ and Heng Fan$^{4}$}
\address{$^1$ Institute of Mathematics, Academy of Mathematics and Systems Science, Chinese Academy of Sciences, Beijing, 100190,China}
\address{$^2$ School of Mathematical Sciences, University of Chinese Academy of Sciences, Beijing, 100190,China}
\address{$^3$ NCMIS, MDIS, Academy of Mathematics and Systems Science, Chinese Academy of Sciences, Beijing, 100190,China}
\address{$^4$ Institute of Physics, Chinese Academy of Sciences, Beijing 100190, China.}
\address{$^5$ School of Physical Sciences, University of Chinese Academy of Sciences, Beijing 100190, China}
\address{$^\dag$ Author to whom any correspondence should be addressed.}
\ead{shangyun@amss.ac.cn}

\vspace{10pt}

\begin{abstract}
In large-scale quantum networks, quantum repeaters provide an efficient method to distribute entangled states among selected nodes for realizing long-distance and complicated quantum communications. However, extending quantum repeater protocols to high-dimensional quantum states in existing experiments is not easy. Owing to the feasible physical implementations of quantum walks, we proposed various basic modules applicable to quantum repeaters for distributing high-dimensional entangled states via quantum walks, including $d$-dimensional Bell states and multi-particle $d$-dimensional GHZ states. Furthermore, based on the above schemes, we provided a high-dimensional entanglement distribution scheme for arbitrary quantum tree networks. By searching for a Steiner tree in a quantum network, we can achieve high-dimensional entanglement distributions over an arbitrary quantum network. We constructed a quantum fractal network based on $d$-dimensional GHZ states and analyzed the quantum transport properties of continuous quantum walks in the network. Compared with the continuous quantum walk on the Sierpinski gasket, the quantum walk on the new fractal network spreads more widely within the same time frame. Finally, we conducted five experiments to implement various basic modules of 2-party or 3-party entanglement distribution schemes in a superconducting quantum processor. Our study can serve as a building block for constructing large and complex quantum networks.
\end{abstract}

\vspace{2pc}

\maketitle


%
%
%
%
\section{Introduction}
Quantum networks hold great potential for realizing various quantum technologies, including secure communication schemes, distributed quantum computing, and metrological applications \cite{62,63,64}. However, constructing large-scale quantum networks to realize long-distance quantum communication is a challenging task in practical quantum information technology \cite{1}. As a powerful and efficient quantum resource, entanglement \cite{2} is the cornerstone of many quantum communication and quantum computing protocols \cite{3,4,5,6}. Many theories and methods related to entangled state generation have been proposed, such as the necessary and sufficient conditions for quantum correlation creation \cite{69} and the entanglement dynamics of quantum systems \cite{70,71}. The quantum network's foundation relies on entangled distributions between nodes; however, creating such distributions over long distances at high data rates remains challenging owing to quantum information degradation during transmission. In practice, quantum repeaters are designed to avoid exponential decay in the success rate or fidelity of quantum information or entangled states by sending light through lossy quantum channels for processing \cite{8,9,10}. Quantum repeaters offer a potential solution by amplifying quantum signals while adhering to the principles of quantum physics. The core idea of a quantum repeater is the entanglement swapping technique, which operates by consolidating multiple short-distance entanglements to distribute a unified long-distance entanglement.

The experimental implementation of quantum repeaters has witnessed significant advancements in recent years \cite{14}. Quantum repeaters mainly have two frameworks: one with quantum memory and the other without \cite{15,22}. The DLCZ scheme \cite{15}, a well-known quantum repeater protocol, uses collective spin excitation in an atomic ensemble \cite{16} to provide the required quantum memory and the heralded entanglement connection used to boost the scaling of efficiency through memory enhancement. In addition to using the ensemble system as quantum memory, there are also single atomic ions or diamond defect spins \cite{17,18,19,20,21}. Another famous experimental scheme proposes all-photonic quantum repeaters to avoid the need for quantum memory by exploiting the graph state in repeater nodes \cite{22}. Quantum repeater protocols have been implemented in several physical systems. However, owing to the challenges of existing experimental techniques, such as performing deterministic high-dimensional multi-particle GHZ state measurements \cite{65}, the swapped entangled states are all 2-dimensional two-particle quantum states \cite{24}, and extending to high-dimensional multi-particle quantum states in existing experiments is not easy. Compared to conventional two-level systems, high-dimensional states can offer extended possibilities such as both higher capacity and noise resilience in quantum communications \cite{25,26}, larger violation of Bell inequality \cite{27}, and more efficient quantum simulation \cite{28} and computation \cite{29}. Although there are many existing studies on qubit quantum repeaters, less attention has been paid to qudit quantum repeaters for quantum entanglement and long-distance distribution of information.

Quantum walks, as a framework combining quantum theory and classical random walk \cite{30,31,32}, play an essential role in quantum information processing, especially for 2-dimensional and high-dimensional quantum teleportation \cite{33} and entangled state generation \cite{34}. As a universal quantum computing model \cite{66,67}, quantum walks can be experimentally realized in many physical systems, including trapped ions \cite{35,36}, nuclear magnetic resonance \cite{37}, photonics \cite{38,39}, neutral atoms \cite{40}, Bose-Einstein condensates \cite{41,42}, cavity quantum electrodynamics \cite{43}, and superconducting qubits \cite{44}. Some of these can be extend to high-dimensional quantum walks \cite{36,41}. Because of the successful physical implementations and entanglement generations of quantum walks, quantum walks have significant potential in the implementation of quantum repeater networks \cite{68}. 

In this paper, based on a quantum walk with coins \cite{45}, we propose a series of quantum repeater protocols to distribute high-dimensional entangled Bell states and GHZ states to avoid high-dimensional Bell state and multi-particle GHZ state measurements. We used them to achieve entangled distributions in quantum tree networks. We then proposed a high-dimensional entanglement distribution protocol over an arbitrary quantum network by finding a Steiner tree within it. The high-dimensional entanglement distribution protocol provides additional implementation frameworks for high-dimensional quantum communications. Simultaneously, we applied our entanglement distribution protocol to design a quantum fractal network based on the Sierpinski gasket as an application. In addition, we analyzed some properties of the new quantum fractal network, such as the quantum transport properties of continuous quantum walks on the network. Compared with the Sierpinski gasket, the new fractal network makes the quantum walk on it spread more widely within the same time frame. Finally, we present some experiment implementations of basic modules on a superconducting quantum processor. To our knowledge, we are the first to experimentally implement the distribution of an entangled state to two parties using two GHZ states and three parties using three Bell or GHZ states.

The remainder of this paper is organized as follows. First, in Section 2, we briefly introduce the model of quantum repeaters and quantum walks with coins. In Section 3, we provide theoretical schemes based on quantum walks for quantum repeaters that can distribute $d$-dimensional quantum entangled states using the Bell and GHZ states with arbitrary particle numbers. In Section 4, we present our entanglement distribution scheme for arbitrary quantum networks according to the above schemes. In Section 5, we describe the construction of a quantum fractal network based on $d$-dimensional GHZ states as an application. We present five experiments to implement various basic modules of entanglement distribution schemes in Section 6. Finally, we provide a summary and outlook in Section 7.

\section{Preliminary}

\subsection{Quantum repeater model.}

\begin{figure}[htbp]
\centering
\subfigure[]{
\begin{minipage}[t]{0.9\linewidth}
\centering
\includegraphics[width=0.9\textwidth]{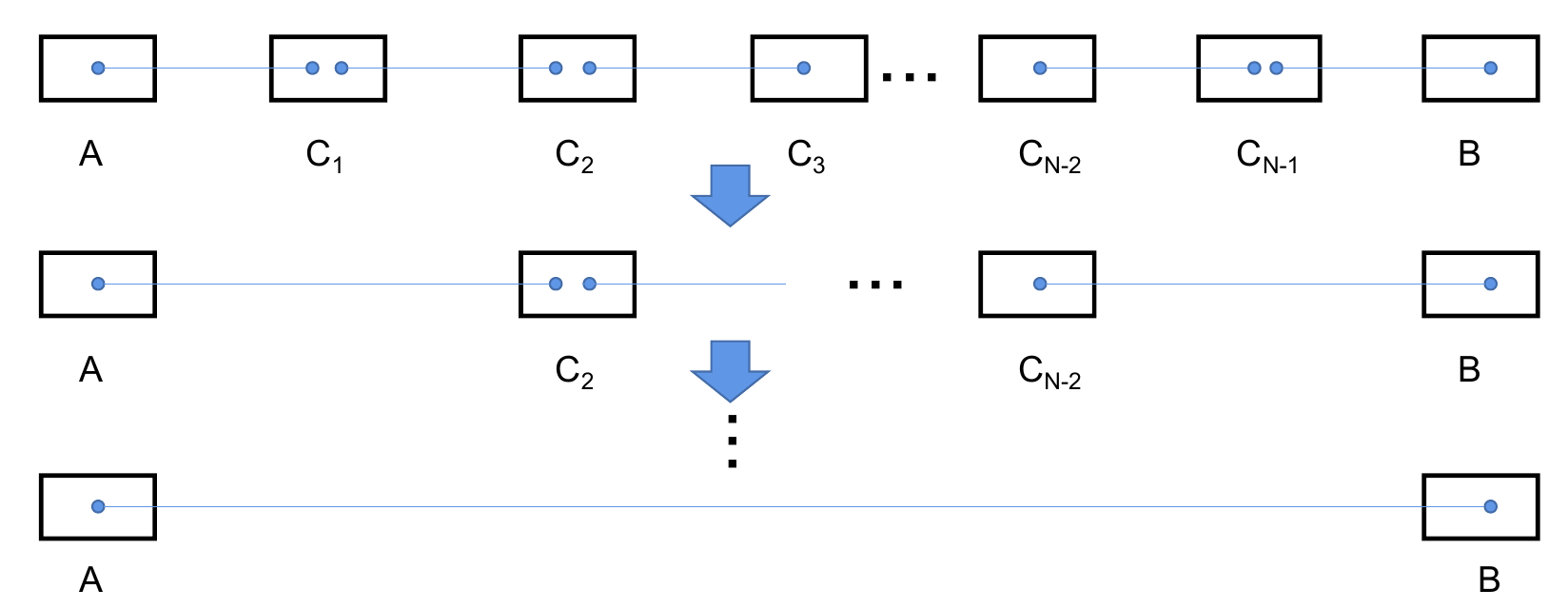}
\label{fig:2}
\end{minipage}
}

\subfigure[]{
\begin{minipage}[t]{0.45\linewidth}
\centering
\includegraphics[width=0.9\textwidth]{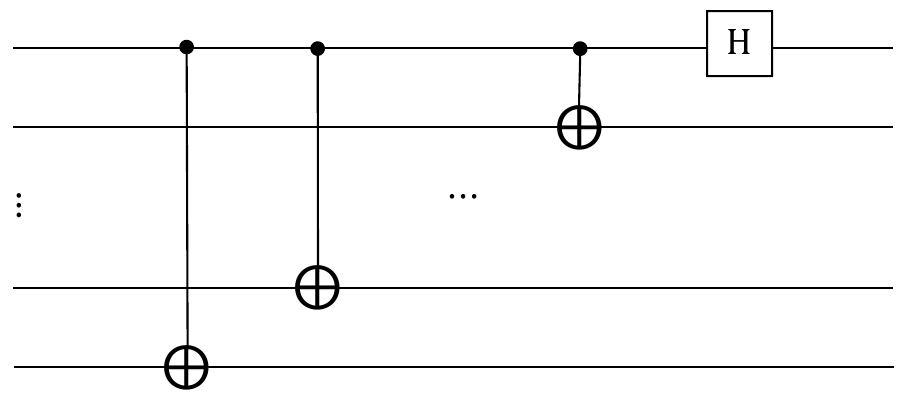}
\label{fig:s1}
\end{minipage}
}
\subfigure[]{
\begin{minipage}[t]{0.45\linewidth}
\centering
\includegraphics[width=0.9\textwidth]{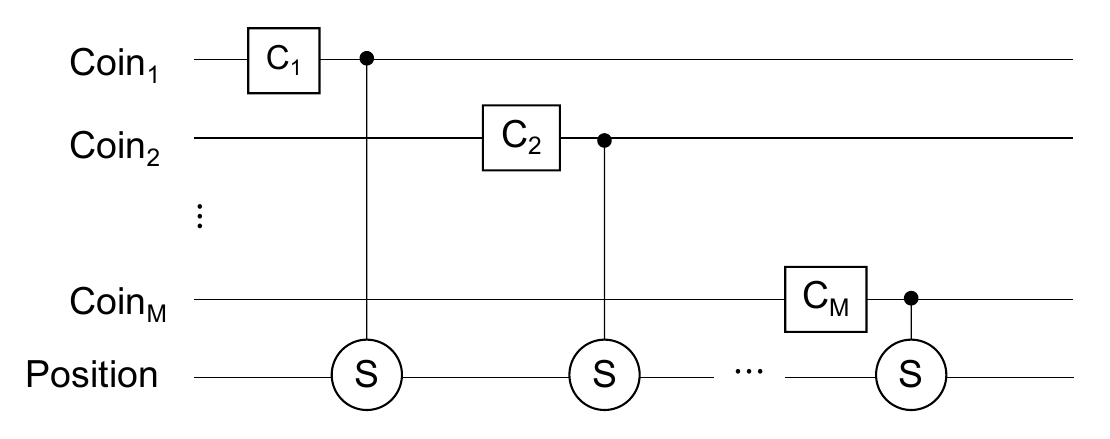}
\label{fig:3}
\end{minipage}
}
\centering
\caption{(a) The model of the quantum repeater. Black squares represent nodes labeled as $A, C_1,\cdots, C_{N-1}, B$, and blue line segments depict entangled states between two nodes. After rounds of connection and purification operations, entangled states will be distributed between two distant nodes, ultimately resulting in an entangled state between nodes $A$ and $B$. (b) The circuit represents a measuring device for N-particle maximally entangled states. (c) The circuit schematic diagram of quantum walk with $M$ coins. The first $M$ horizontal lines represent the coin states, and the last line represents the position state. $C_i$ is the coin operator acting on the $i$-th coin space. $S$ is the shift operator acting on the combination space of the position and the coin space.}
\end{figure}

Quantum repeaters play a crucial role in establishing long-distance quantum communication. According to the quantum repeater scheme, the communication channel is divided into $N$ segments with connection points or auxiliary nodes interspersed. Prepare $N$ elementary entangled states such as EPR pairs between nodes $A$ and $C_1$, $C_1$ and $C_2$,...,$C_{N-1}$ and $B$, as shown in \Cref{fig:2}. To distribute new entangled states between more distant nodes, every pair of neighboring elementary entangled states is connected at intermediary nodes using methods such as Bell state measurements or GHZ state measurements as shown in \Cref{fig:s1}. Except for EPR pairs, elementary entangled states encompass $d$-dimensional Bell or GHZ states, which allow for high-dimensional quantum communication with an increased message capacity. GHZ states exhibit enhanced robustness under local decoherence \cite{46}. After establishing entangled states between distant nodes, entanglement purification protocols are utilized to enhance the fidelity of the new entangled states. This study primarily focused on the entanglement distribution process of a quantum repeater through quantum walks with coins.  

\subsection{Quantum walks with multiple coins on complete graphs.}

A quantum walk with $M$ coins on the $d$-complete graph takes place in a compound Hilbert space comprising a position space and $M$ coin spaces, i.e. $\mathcal{H}=\mathcal{H}_{C_1}\otimes\cdots\otimes\mathcal{H}_{C_M}\otimes\mathcal{H}_P$, where $\mathcal{H}_P=span\{|n\rangle: n=0,1,\cdots d-1\}$, and $|n\rangle$ is the position state corresponding to the vertex $n$ on the $d$-completed graph. At each vertex $n$, there are $d$ directed edges with labels $\{0, 1, \cdots d-1\}$ pointing to other vertices. For any coin space, $\mathcal{H}_{C_m}=span\{|a\rangle: a=0,1,\cdots d-1\}, m=1, 2, \cdots, M$, and $|a\rangle$ is the coin state corresponding to the edge $a$. The conditional shift operator acting on the $m$-th coin space is given by 
$S=\sum\limits_{k,j=0}^{d-1}|k\rangle_{C_m}\langle k|\otimes|j-k\rangle_P\langle j|$. Especially, when $d=2$, the $\mathcal{H}_P=span\{|0\rangle, |1\rangle\}$, and $\mathcal{H}_{C_m}=span\{|0\rangle, |1\rangle\}, m=1,2, \cdots, M$. Thus the $m$-th step quantum walk is described as $W_m=S(C_m\otimes I)$, where the coin operator $C_m$ and the identify operator $I$ act on the coin space $\mathcal{H}_{C_m}$ and the position space, respectively. The conditional shift operator $S$ acts on the space ${H}_{C_m}\otimes\mathcal{H}_P$. A circuit diagram of this quantum walk is shown in \Cref{fig:3}. It is worth noting that the conditional shift operator $S^{2-complete}$ of the quantum walk on $2$-complete graph can be easily simulated as the CNOT gate using digital quantum circuit models. For the conditional shift operator $S^{d-complete}$ of the quantum walk on $d$-complete graph, its digital physical implementation is not easy for digital quantum circuits, particularly in the noisy intermediate-scale quantum era \cite{80}, but it can be physically implemented as a high-dimensional generalized CNOT gate by using optical quantum systems \cite{79}. Thus, the quantum walks we employed can be physically realized in existing physical systems, further demonstrating the feasibility of our designed entanglement distribution scheme in physical experiments.  

\begin{figure}[htbp]
\centering
\subfigure[]{
\begin{minipage}[t]{0.45\linewidth}
\centering
\includegraphics[width=0.9\textwidth]{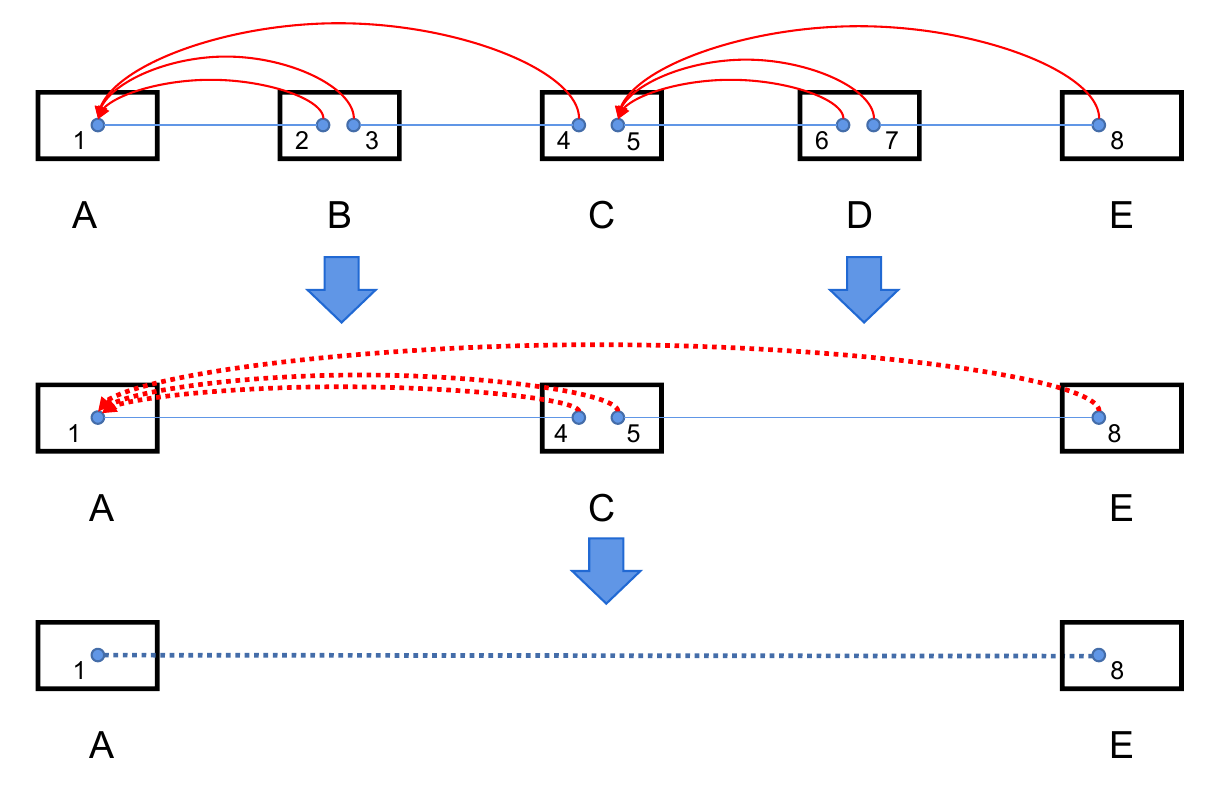}
\label{fig:4(a)}
\end{minipage}
}
\subfigure[]{
\begin{minipage}[t]{0.45\linewidth}
\centering
\includegraphics[width=0.9\textwidth]{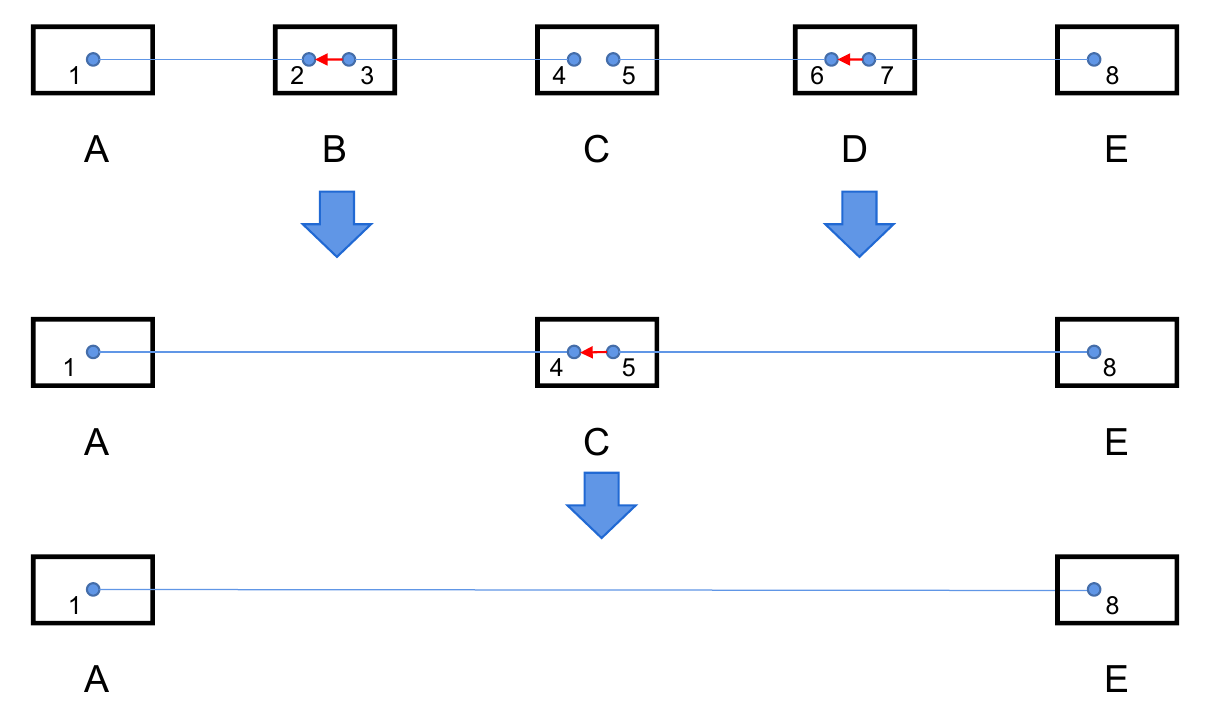}
\label{fig:4(b)}
\end{minipage}
}
\centering
\caption{(a) The schematic diagrams of the entanglement generation schemes proposed by \cite{34}. Black squares represent nodes labeled as $A, B, C, D, E$, blue line segment represents an entangled state between two nodes. The red arrows within a square symbolize quantum walk operations, with their origin points denoting coin states and their destination points corresponding to position states. The red dashed arrow represents the theoretically achievable quantum walk to generate entangled states between nodes $A$ and $E$. However, when the distance between nodes $A$ and $E$ exceeds the limit of the quantum operation reachable distance, the quantum walk between nodes $A$ and $E$ cannot be directly achieved in experiments. Thus, the blue dashed line represents the theoretical entangled state, but it cannot be realized in experiments, when the distance between nodes $A$ and $E$ exceeds the limit of the quantum operation reachable distance. (b) The schematic diagrams of the local operations that apply to the framework of quantum repeaters. We can see that as the distance between the nodes of the entangled distribution increases, all the quantum operations are local. Therefore, the quantum repeater framework has no fundamental limitation on the achieved distance.} 
\end{figure}

Although \cite{34} proposed some schemes to prepare entangled states using quantum walks, it is not feasible to implement an entanglement distribution using their entanglement generation schemes. If the schemes are used to distribute distant entangled states, we must implement operations across nodes as shown in \Cref{fig:4(a)}. Specifically, in the $n$-th step of the entanglement distribution,  if the distance between each pair of adjacent nodes is $l$, we must implement some operations between two nodes with a distance of $2^nl$ or $2^{n-1}l$, which will be long-distance operations as $n$ increases. Finally, we still need to perform long-distance quantum operations between the nodes desired for distributed entanglement, such as nodes $A$ and $E$ in \Cref{fig:4(a)}. Therefore, this scheme is a point-to-point quantum communication scheme \cite{81} instead of a quantum repeater scheme. Although the range of experimentally implementing long-distance quantum operations may be continually expanding, for example, the farthest distance of a nonlocal CNOT gate implemented experimentally is $7$ km \cite{82}, there are fundamental limitations on the achievable distance for point-to-point quantum communication schemes for various quantum channels such as bosonic lossy channels and quantum-limited amplifiers \cite{83}. Thus, when the distance between nodes that want to distribute entanglement exceeds the fundamental limit of the achievable distance of quantum operations, the entanglement generation scheme in reference \cite{34} is not feasible for entanglement distribution. Therefore, to achieve an entanglement distribution using quantum repeater schemes that have no fundamental limitations on the achievable distance, all quantum operations that can be adopted should be local, that is, in a node. After the implementation of these operations, the distributed quantum states should be entangled states of the same type as the initial entangled state, allowing the entanglement distribution to proceed iteratively, as shown in \Cref{fig:4(b)}. Thus, all the operations of the entanglement distribution schemes we propose are local operations that fit within the framework of quantum repeaters.

\section{Basic modules of entanglement distribution}

\subsection{D-dimensional quantum states}

In this section, we discuss the basic modules required for the distribution of $d$-dimensional entangled states implemented by local quantum walks with coins on a $d$-complete graph. We focus on the basic modules constructed from $d$-dimensional Bell and GHZ states, as shown in \Cref{fig:7}.

\subsubsection{D-dimensional Bell states}

Generalized $d$-dimensional Bell states, a basis of maximally entangled $d$-dimensional bipartite states, can be described as
\begin{eqnarray}
    |\psi_{m,n}\rangle=\frac{1}{\sqrt{d}}\sum\limits_{i=0}^{d-1}\omega^{mi}|i, i-n\rangle,
\end{eqnarray}
where $\omega=e^{\frac{2\pi i}{d}}$. We can distribute a $d$-dimensional $M$-particles GHZ state through quantum walk operations in one node to $M+1$ nodes entangled with the node by $M$ $d$-dimensional Bell states, as shown in \Cref{fig:7(c)}. Here we only consider the $d$-dimensional Bell state $|\psi_{0,0}\rangle=\frac{1}{\sqrt{d}}\sum\limits_{i=0}^{d-1}|i, i\rangle$, since $|\psi_{0,0}\rangle$ can be transformed from the generalized $d$-dimensional Bell state $|\psi_{m,n}\rangle$ by acting on the first qudit with $U_{m,n}=\sum\limits_{i=0}^{d-1}\omega^{-mi}|i-n\rangle \langle i|$.
Thus the initial state is
\begin{eqnarray}
|\Psi(0)\rangle=\bigotimes\limits_{k=1}^{M-1}\big(\frac{1}{\sqrt{d}}\sum\limits_{i_k=0}^{d-1}|i_k, i_k\rangle_{2k-1,2k}\big)\otimes\frac{1}{\sqrt{d}}\sum\limits_{j=0}^{d-1}|j, j\rangle_{2M-1,2M}.
\end{eqnarray}
Take particles $2k (k= 1,2,\cdots, M-1)$ as the coin state at the $k-th$ step of the quantum walk and particle $2M-1$ as the position state in the yellow node. Set the coin operator as $C=F$, and the conditional shift operator as $S=\sum\limits_{k,j=0}^{d-1}|k\rangle\langle k|\otimes|j-k\rangle\langle j|$. After performing $M-1$ steps of the quantum walk, the quantum state becomes 
\begin{align}
|\Psi(M-1)\rangle=\bigotimes\limits_{k=1}^{M-1}\big(\frac{1}{d}\sum\limits_{i_k,p_k=0}^{d-1}\omega^{i_kp_k}|i_k, p_k\rangle_{2k-1,2k}\big)\otimes\frac{1}{\sqrt{d}}\sum\limits_{j=0}^{d-1}|j-\sum\limits_{k=1}^{M-1}p_k, j\rangle_{2M-1,2M}.
\end{align}
Then, we measure with the basis $\{|\widetilde{k}\rangle: k=0,\cdots,d-1\}$ and $\{|k\rangle: k=0,\cdots,d-1\}$ on the coin and position states, respectively, where the measurement results are recorded as $\widetilde{q_{k0}}$ and $u_{0}$. We perform the quantum Fourier inverse transformation on particle $2M$, and the remaining quantum state is $\frac{1}{\sqrt{d}}\sum\limits_{r=0}^{d-1}\omega^{-ru_0}\bigotimes\limits_{k=1}^{M-1}|r+\widetilde{q_{k0}}\rangle_{2k-1}\otimes|r\rangle_{2M}$, which is a general $d$-dimensional $M$-particles GHZ state distributed in the $M$ nodes. If particle $2M$ is also in the yellow node, as shown in \Cref{fig:7(d)}, then we can perform the above operations to distribute a $d$-dimensional $M$-particles GHZ state between the yellow node and other $M-1$ nodes.

\subsubsection{D-dimensional GHZ states}

We can distribute a $d$-dimensional GHZ state through quantum walk operations in the yellow node to nodes entangled with the yellow node by 2 $d$-dimensional GHZ states.

\begin{figure}[htbp]
\centering
\subfigure[]{
\begin{minipage}[t]{0.45\linewidth}
\centering
\includegraphics[width=0.9\textwidth]{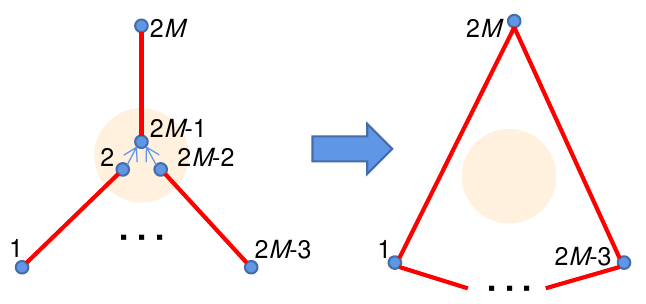}
\label{fig:7(c)}
\end{minipage}
}
\subfigure[]{
\begin{minipage}[t]{0.45\linewidth}
\centering
\includegraphics[width=0.9\textwidth]{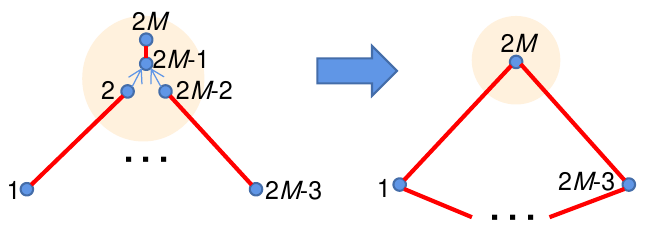}
\label{fig:7(d)}
\end{minipage}
}

\subfigure[]{
\begin{minipage}[t]{0.55\linewidth}
\centering
\includegraphics[width=0.9\textwidth]{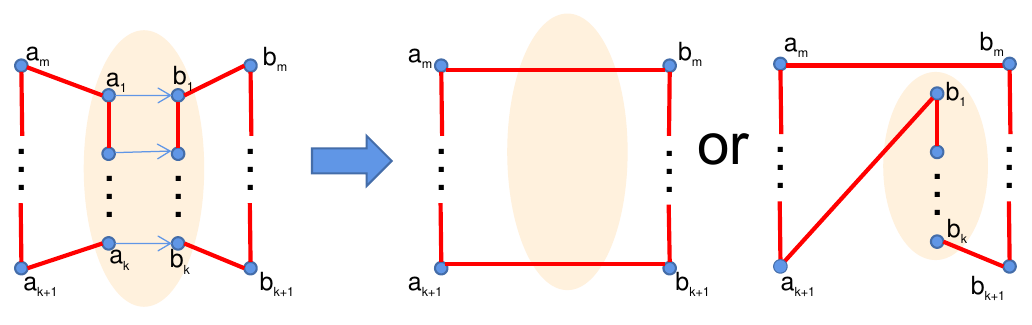}
\label{fig:7(a)}
\end{minipage}
}
\subfigure[]{
\begin{minipage}[t]{0.4\linewidth}
\centering
\includegraphics[width=0.9\textwidth]{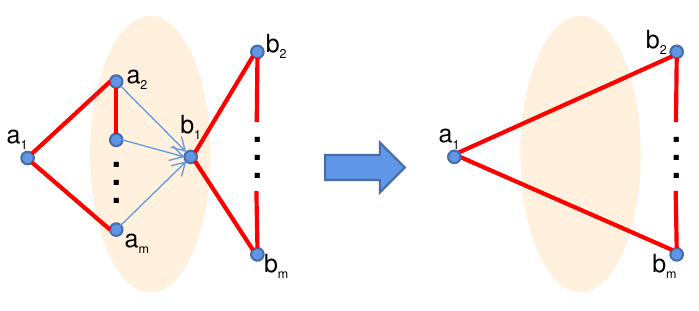}
\label{fig:7(b)}
\end{minipage}
}

\centering

\caption{(a)-(b)  Schematic diagrams for distributing a $d$-dimensional multi-particle GHZ states by multiple d-dimensional Bell states. Red line segments and polygons represent $d$-dimensional Bell states and GHZ states with vertexes being as qudits. Arrows in the yellow node represent quantum walk operations with start points and target points being coin states and position states, respectively. (c)-(d)  Schematic diagrams for distributing a $d$-dimensional multi-particle GHZ states using two multi-particle GHZ states.}
\label{fig:7}
\end{figure}

We can perform parallel quantum walks with one coin to distribute a $d$-dimensional GHZ state as shown in \Cref{fig:7(a)}. Let the initial state be
\begin{eqnarray}
|\Psi(0)\rangle&=\frac{1}{\sqrt{d}}\sum\limits_{i=0}^{d-1}|i, i,\cdots,i\rangle_{a_1,a_2,\cdots,a_m}\otimes\frac{1}{\sqrt{d}}\sum\limits_{j=0}^{d-1}|j, j,\cdots,j\rangle_{b_1,b_2,\cdots,b_n}.
\end{eqnarray}
We let the particles $a_1, \cdots, a_k$ be the coin states and the particle $b_1, \cdots, b_k$ be the corresponding position states to perform $k$ parallel quantum walks. Then, the quantum state becomes
\begin{align}
|\Psi(1)\rangle&=\frac{1}{\sqrt{d}}\sum\limits_{i=0}^{d-1}|i, i,\cdots,i\rangle_{a_1,a_2,\cdots,a_k, a_{k+1},\cdots,a_m}\nonumber\\
&\otimes|j-i, j-i,\cdots,j-i,j,\cdots,j \rangle_{b_1,b_2,\cdots,b_k, b_{k+1},\cdots,b_n}.
\end{align}
Let's measure with the basis $\{|\widetilde{k}\rangle: k=0,\cdots,d-1\}$ and $\{|k\rangle: k=0,\cdots,d-1\}$ on the particles $a_1, \cdots, a_k$ and $b_1, \cdots, b_k$, respectively. If the measurement results of the particles $a_1, \cdots, a_k$ are $\widetilde{p_1},\cdots,\widetilde{p_k}$, respectively. And the measurement results of the particles $b_1, \cdots, b_k$ are $u_0$, then the entanglement state of leftover particles is 
\begin{align}
|\Psi\rangle=\frac{1}{\sqrt{d}}\sum\limits_{i=0}^{d-1}\omega^{i(-\widetilde{p_1}-\cdots-\widetilde{p_k})}|i,\cdots,i\rangle_{a_{k+1},\cdots,a_m}\otimes|i+u_0,\cdots,i+u_0\rangle_{b_{k+1},\cdots,b_n},
\end{align}
which can be recovered to a $d$-dimensional GHZ state by local unitary operations according to the measurement results. If we only measure the particles $b_1,\cdots, b_k$, we can obtain a GHZ state distributed between yellow node and other entangled nodes.

We can also adopt a quantum walk with multiple coins to distribute a $d$-dimensional GHZ state as shown in \Cref{fig:7(b)}. Let the initial state be 
\begin{eqnarray}
|\Psi(0)\rangle&=\frac{1}{\sqrt{d}}\sum\limits_{i=0}^{d-1}|i, i,\cdots,i\rangle_{a_1,a_2,\cdots,a_m}\otimes\frac{1}{\sqrt{d}}\sum\limits_{j=0}^{d-1}|j, j,\cdots,j\rangle_{b_1,b_2,\cdots,b_n},
\end{eqnarray}
where the particles $a_2, \cdots, a_m$ and the particle $b_1$ are the coin states and the position state, respectively. The coin operator $C_1=\cdots=C_{m-1}=F$, and the conditional shift operator $S=\sum\limits_{k,j=0}^{d-1}|k\rangle\langle k|\otimes|j-k\rangle\langle j|$. After $m-1$ steps of the quantum walk, the quantum state is
\begin{align}
|\Psi(1)\rangle&=\frac{1}{\sqrt{d^{m+1}}}\sum\limits_{i,j,k_1,\cdots,k_{m-1}=0}^{d-1}\omega^{i\sum\limits_{l=1}^{m-1}k_l}|i, k_1,\cdots,k_{m-1}\rangle_{a_1,a_2,\cdots,a_m}\nonumber\\
&\otimes|j-\sum\limits_{l=1}^{m-1}k_l, j,\cdots,j\rangle_{b_1,b_2,\cdots,b_n}\nonumber\\
&=\frac{1}{d^m}\sum\limits_{i,j,k_1,\cdots,k_{m-1}=0}^{d-1}\sum\limits_{p_1,\cdots,p_{m-1}=0}^{d-1}\omega^{i\sum\limits_{l=1}^{m-1}k_l-\sum\limits_{l=1}^{m-1}k_lp_l}\times|i, \widetilde{p_1},\cdots,\widetilde{p_{m-1}}\rangle_{a_1,a_2,\cdots,a_m}\nonumber\\
&\otimes|j-\sum\limits_{l=1}^{m-1}k_l, j,\cdots,j\rangle_{b_1,b_2,\cdots,b_n}.
\end{align}
Next, we perform the inverse quantum Fourier transformation on the particle $a_1$, and let $\sum\limits_{l=1}^{m-1}k_l=a$, then the state is 
\begin{align}
|\Psi(2)\rangle=\frac{1}{d\sqrt{d}}\sum\limits_{a,j,p_1=0}^{d-1}\omega^{-ap_1}|a, \widetilde{p_1},\cdots,\widetilde{p_1}\rangle_{a_1,a_2,\cdots,a_m}\otimes|j-a, j,\cdots,j\rangle_{b_1,b_2,\cdots,b_n}.
\end{align}
Measure with the basis $\{|\widetilde{k}\rangle: k=0,\cdots,d-1\}$ in the particles $a_2,\cdots,a_m$ and $\{|k\rangle: k=0,\cdots,d-1\}$ in the particle $b_1$, respectively. If the measurement results of particles $a_2,\cdots,a_m$ are $\widetilde{p_0}$, and the measurement result of particle $b_1$ is $u_0$ respectively, then the entanglement state of the leftover particles is 
\begin{eqnarray}
|\Psi(3)\rangle=\frac{1}{\sqrt{d}}\sum\limits_{a=0}^{d-1}\omega^{-a\widetilde{p_0}}|a, a+u_0,\cdots, a+u_0\rangle_{a_1,b_2,\cdots,b_n},
\end{eqnarray}
which can be recovered to a $d$-dimensional GHZ state through local unitary operators easily.

\begin{figure}[htbp]
\centering
\subfigure[]{
\begin{minipage}[t]{0.6\linewidth}
\centering
\includegraphics[width=0.9\textwidth]{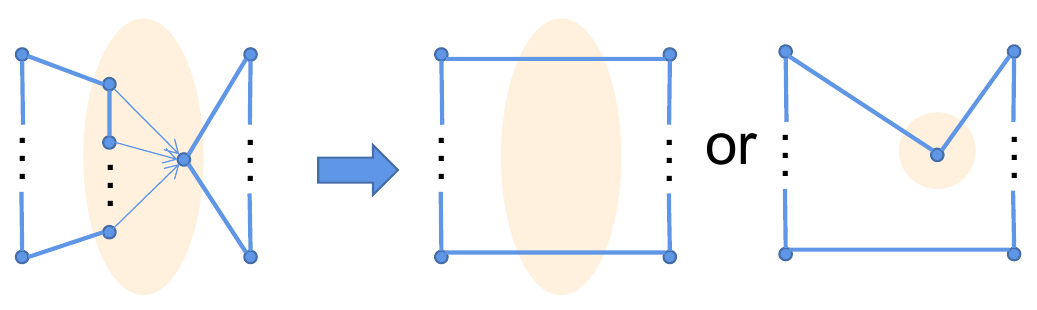}
\label{fig:6(a)}
\end{minipage}
}
\subfigure[]{
\begin{minipage}[t]{0.35\linewidth}
\centering
\includegraphics[width=0.9\textwidth]{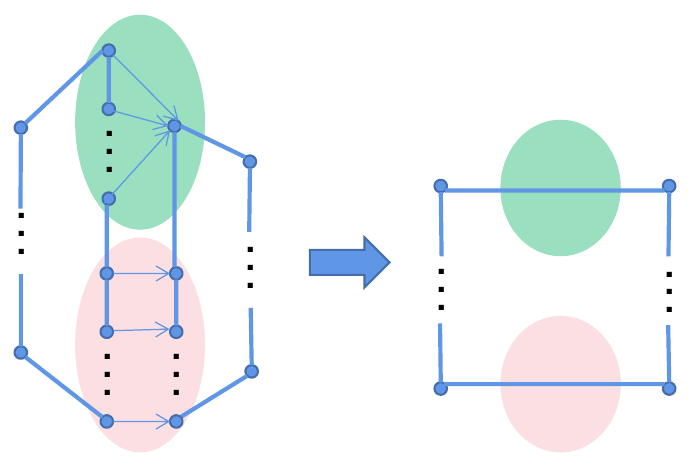}
\label{fig:6(c)}
\end{minipage}
} 
\caption{(a) The blue polygons represent $2$-dimensional GHZ states, where the vertices represent qubits. After performing quantum walk and measurement operations, we distribute a GHZ state between the nodes outside the yellow region or between other nodes and the yellow one. (b) The schematic diagram illustrates the operational process of simultaneously employing the aforementioned two methods to distribute a GHZ state.}
\end{figure}

Compared to the high-dimensional module shown in \Cref{fig:7(b)}, the corresponding 2-dimensional module is more flexible. We can choose an arbitrary number of particles as coin states to perform multi-coin quantum walks as shown in \Cref{fig:6(a)}. Let the initial state be 
\begin{align}
|\Psi(0)\rangle=\frac{|0\cdots00\cdots0\rangle+|1\cdots11\cdots1\rangle}{\sqrt{2}}_{a_1,\cdots, a_k, a_{k+1},\cdots, a_m}\otimes\frac{|00\cdots0\rangle+|11\cdots1\rangle}{\sqrt{2}}_{b_1,b_2,\cdots, b_n}
\end{align}
We can let the particles $a_1,a_2, \cdots, a_k$ be the coin states and particle $b_1$ be the position state. The coin operators were set as $C_1=X, C_2=C_3=\cdots=C_k=H$, and the conditional shift operator $S$ was set as the CNOT gate. 
Then, the quantum state after the first step of the quantum walk is 
\begin{align}
|\Psi(1)\rangle&=\frac{1}{2}(|10\cdots00\cdots010\cdots0\rangle+|10\cdots00\cdots001\cdots1\rangle\nonumber\\
&+|01\cdots11\cdots100\cdots0\rangle+|01\cdots11\cdots111\cdots1\rangle)_{a_1,a_2, \cdots, a_k,a_{k+1},\cdots, a_m, b_1,b_2,\cdots, b_n}.
\end{align}
After the remaining $k-1$ steps of the quantum walk, the quantum state is 
\begin{align}
|\Psi(k)\rangle&=\frac{1}{(\sqrt{2})^{k+1}}\sum\limits_{x_2,\cdots x_k=0}^{1}(|1x_2\cdots x_k0\cdots0\overline{y}0\cdots0\rangle+|1x_2\cdots x_k0\cdots0y1\cdots1\rangle\nonumber\\
&+(-1)^{y}|0x_2\cdots x_k1\cdots1y0\cdots0\rangle+(-1)^{y}|0x_2\cdots x_k1\cdots1\overline{y}1\cdots1\rangle) 
\end{align}
where $y=(x_2+\cdots+x_k)$ mod 2.
Then, we take a measurement with the basis $\{|+\rangle, |-\rangle\}$ in the particle $a_1$ and the basis $\{|0\rangle, |1\rangle\}$ in the particles $a_2,\cdots, a_k$ and $b_1$. Then the entanglement states of the particles $a_{k+1}, \cdots, a_m$ and $b_2, \cdots, b_n$ can be recovered to the state $\frac{|0\cdots0\rangle+|1\cdots1\rangle}{\sqrt{2}}$ by some local unitary operations according to the measurement results. If we only measure the particles $a_2,\cdots, a_k$ and $b_1$, we also can obtain a GHZ state entangles the yellow node and other nodes. 

More generally, we can distribute GHZ states with $m+n-q, 2\leq q \leq m+n-2$ qubits using the above methods simultaneously, as shown in \Cref{fig:6(c)}. Specifically, for the initial state,     
\begin{align}
|\Psi(0)\rangle&=\frac{|0\cdots00\cdots0\rangle+|1\cdots11\cdots1\rangle}{\sqrt{2}}_{a_1,\cdots,a_k,a_{k+1},\cdots, a_m}\nonumber\\
&\otimes\frac{|0\cdots00\cdots0\rangle+|1\cdots11\cdots1\rangle}{\sqrt{2}}_{b_1,\cdots, b_l,b_{l+1},\cdots, b_n},
\end{align}
without losing generality, we can assume $k>l\geq2$ and $k+l=q$. We perform parallel quantum walk with one coin in the particles $a_1,\cdots,a_{l-1}$ and $b_1,\cdots,b_{l-1}$, and quantum walk with multiple coins in the particles $a_l,\cdots,a_k$ and $b_l$. Thus, the initial quantum state becomes
\begin{align}
|\Psi(1)\rangle&=\frac{1}{(\sqrt{2})^{k-l}}\sum\limits_{x_{l+1},\cdots,x_k=0}^{1}(|1\cdots11x_{l+1}\cdots x_k0\cdots0\rangle\nonumber\\
&\otimes|1\cdots1\overline{y}0\cdots0\rangle+|1\cdots11x_{l+1}\cdots x_k0\cdots0\rangle\nonumber\\
&\otimes|0\cdots0y1\cdots1\rangle+(-1)^y|0\cdots00x_{l+1}\cdots x_k1\cdots1\rangle\nonumber\\
&\otimes|0\cdots0y0\cdots0\rangle+(-1)^y|0\cdots00x_{l+1}\cdots x_k1\cdots1\rangle\nonumber\\
&\otimes|1\cdots1\overline{y}1\cdots1\rangle),&
\end{align}
where $y=(x_{l+1}+\cdots+x_k)$ mod 2. After the measurement and local unitary operators, we obtain a GHZ state with $m+n-q$ qubits. 

By performing quantum walk operations on any number of particles in two GHZ states at an intermediate node, we can orchestrate the distribution of an entangled state with an increased span or a longer distance. This capability enhances the quantum repeater network's potential to create extended-distance entanglement using GHZ states with various particle numbers, thereby diversifying the quantum repeater designs for quantum networks. Moreover, the multi-particle GHZ state proves advantageous in mitigating quantum noise, making it a more effective option for applications in quantum communication.

\section{Entanglement distribution on arbitrary quantum networks}

In this section, we discuss the multi-particle entangled state distribution in an arbitrary quantum network using the basic modules mentioned above. In our quantum network, we can store qudits (or qubits) and perform local operations at each node. Each edge represents an entangled state between two nodes. The basic modules of entanglement distribution are employed in the aforementioned quantum repeater, wherein a multi-particle entangled state is distributed among arbitrarily selected nodes using basic entangled states within the network and quantum walk operations.

\begin{figure}[htbp]
\centering
\subfigure[]{
\begin{minipage}[t]{0.45\linewidth}
\centering
\includegraphics[width=0.8\textwidth]{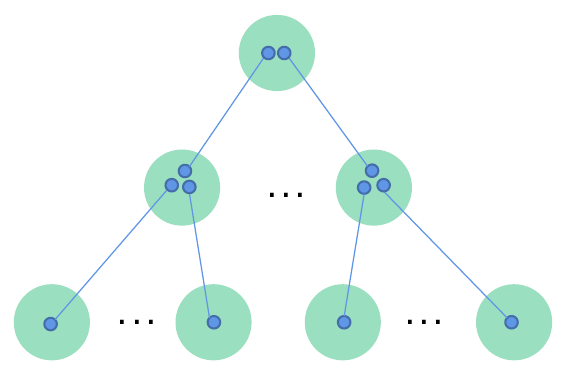}
\label{fig:14(a)}
\end{minipage}%
}%
\subfigure[]{
\begin{minipage}[t]{0.45\linewidth}
\centering
\includegraphics[width=0.8\textwidth]{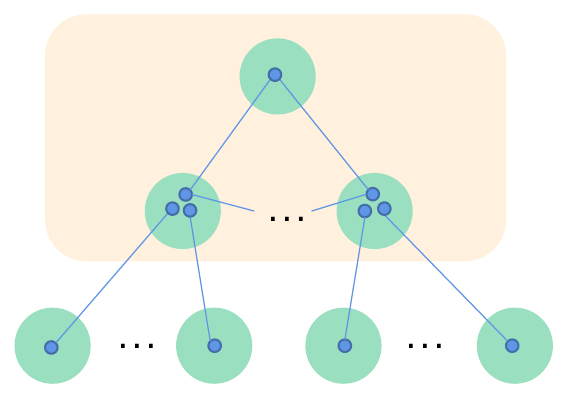}
\label{fig:14(b)}
\end{minipage}%
}%

\subfigure[]{
\begin{minipage}[t]{0.45\linewidth}
\centering
\includegraphics[width=0.8\textwidth]{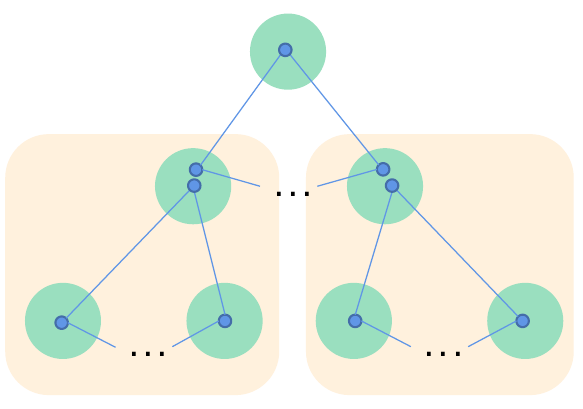}
\label{fig:14(c)}
\end{minipage}
}%
\subfigure[]{
\begin{minipage}[t]{0.45\linewidth}
\centering
\includegraphics[width=0.8\textwidth]{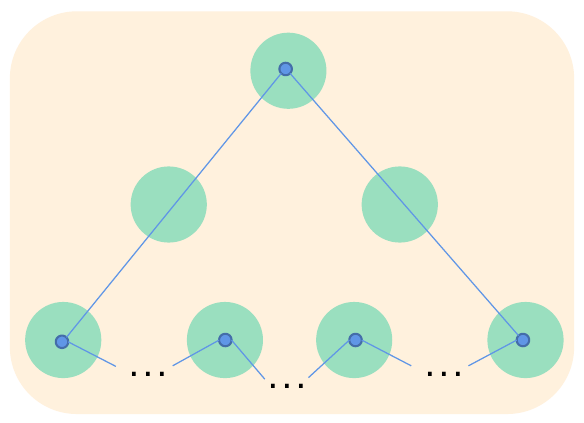}
\label{fig:14(d)}
\end{minipage}
}%

\centering

\caption{The schematic diagram of distributing a GHZ state between the root node and the level 2 nodes. (a) A tree network with the root node and two level nodes. (b) and (c) We perform quantum walks between each parent node and its child nodes to obtain a GHZ state in yellow areas. (d) We perform quantum walks on the two GHZ states in each middle level node to distribute a GHZ state between the root node and the level 2 nodes.}
\label{fig:14}
\end{figure}

We aim to connect the selected nodes in a quantum network through a connected graph with minimal edges, which can be achieved by implementing the Steiner tree algorithm\cite{58}. This approach allows additional points to be added outside of the given nodes, resulting in a connected graph with minimal overhead. The cost consideration involves determining the number of basic entangled states or edges required to generate a connected graph that connects these selected nodes with a minimum edge count, that is, a Steiner tree. We only need to consider the distribution of the entangled state among the selected nodes in the tree network. The next step involved determining the optimal distribution of the entangled state among a specific subset of nodes in the tree network. Once a node is chosen as the root, the remaining nodes can be categorized into different levels based on their distance from the root. It should be noted that all leaf nodes were considered as part of this selected subset. \Cref{fig:14} shows the process of distributing the entangled state between the root node and level 2 nodes. For the three adjacent level nodes, as shown in \Cref{fig:14(a)}, we perform the operation shown in \Cref{fig:7(d)} between each parent and its child nodes to obtain a GHZ state in two adjacent level nodes. Each node can share a GHZ state with its parent and sibling nodes, and it can also share a GHZ state with its child nodes, as shown in \Cref{fig:14(b)} and \Cref{fig:14(c)}. Then, we perform the operations shown in \Cref{fig:7(a)} on the two GHZ states in each node, and we can obtain the GHZ states between the next-closest-level nodes as shown in \Cref{fig:14(d)}. The above operations are repeated to obtain the GHZ state between nodes at arbitrary levels. We can perform the corresponding measurement operations according to whether the node is a selected node to keep or not keep the qubits or qudits in the node. Thus, we can realize GHZ state distribution between chosen nodes in the tree network. 

\begin{figure}[htbp]
\centering
\subfigure[]{
\begin{minipage}[t]{0.3\linewidth}
\centering
\includegraphics[width=1\textwidth]{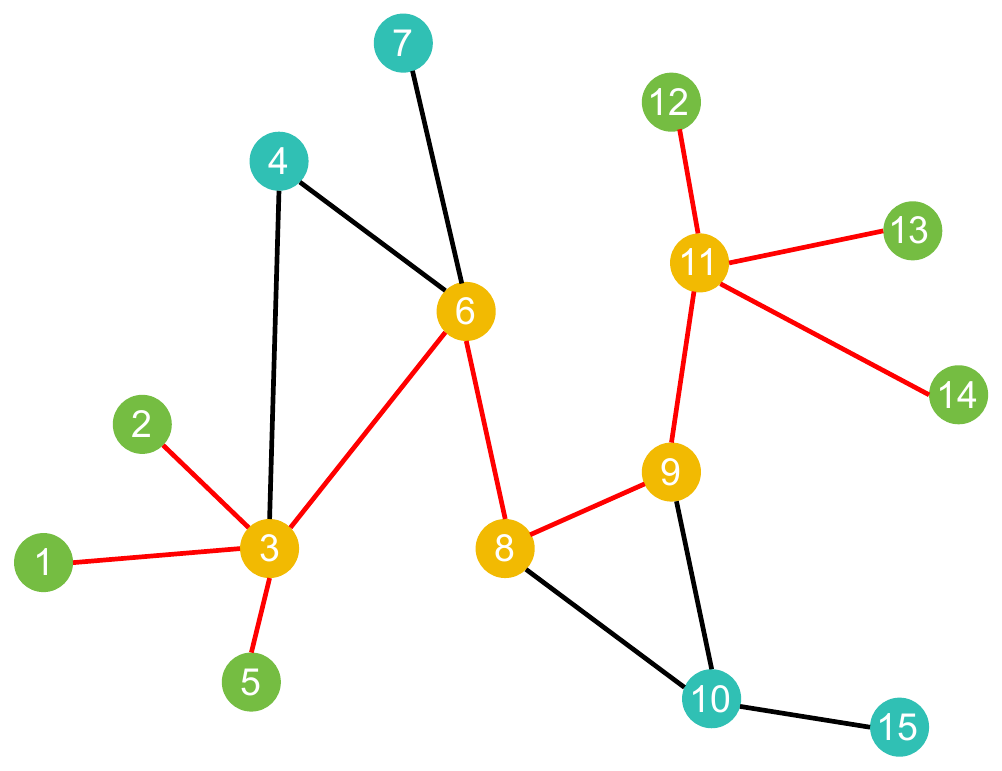}
\label{fig:15(a)}
\end{minipage}%
}%
\subfigure[]{
\begin{minipage}[t]{0.3\linewidth}
\centering
\includegraphics[width=1\textwidth]{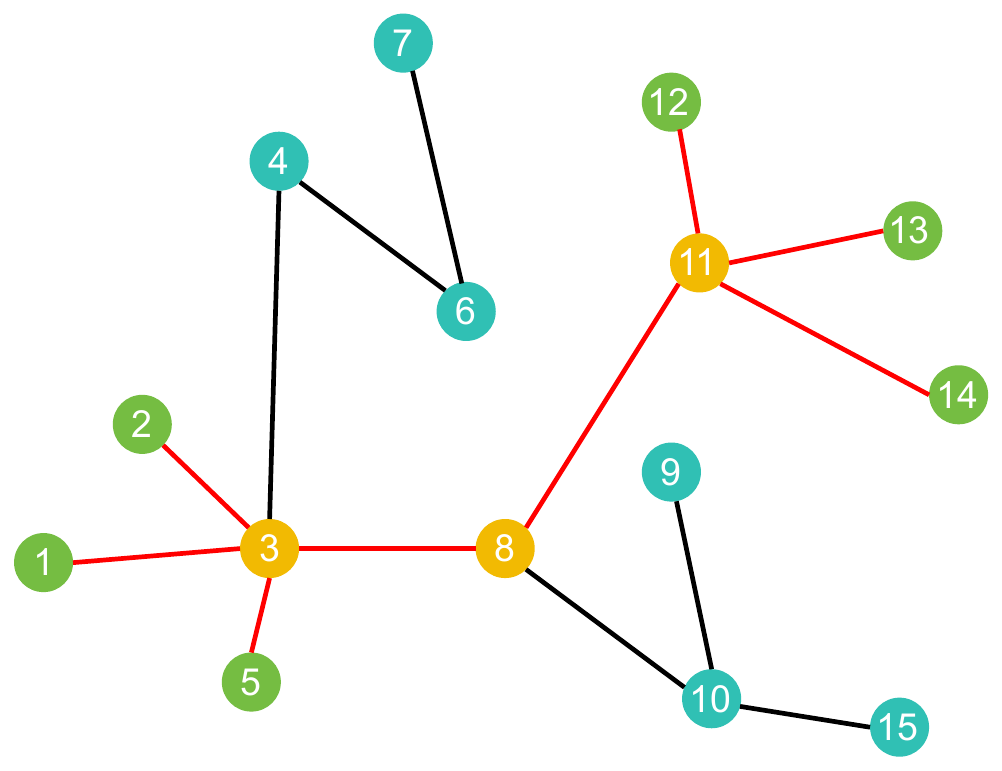}
\label{fig:15(b)}
\end{minipage}%
}%
\subfigure[]{
\begin{minipage}[t]{0.3\linewidth}
\centering
\includegraphics[width=1\textwidth]{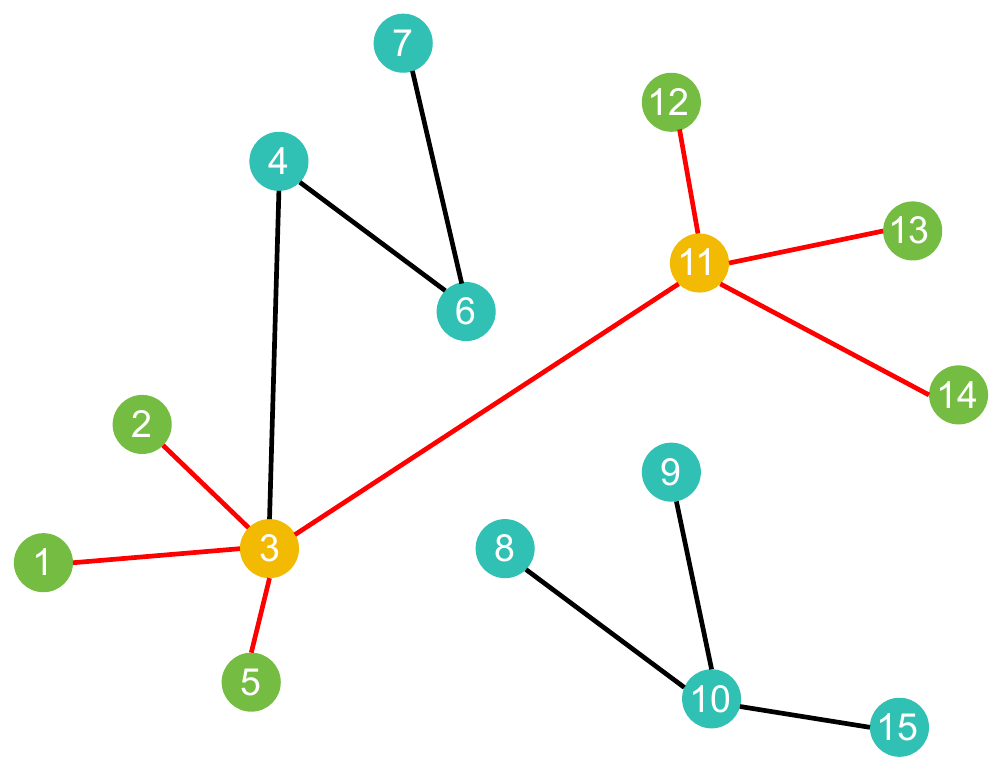}
\label{fig:15(c)}
\end{minipage}%
}%

\subfigure[]{
\begin{minipage}[t]{0.3\linewidth}
\centering
\includegraphics[width=1\textwidth]{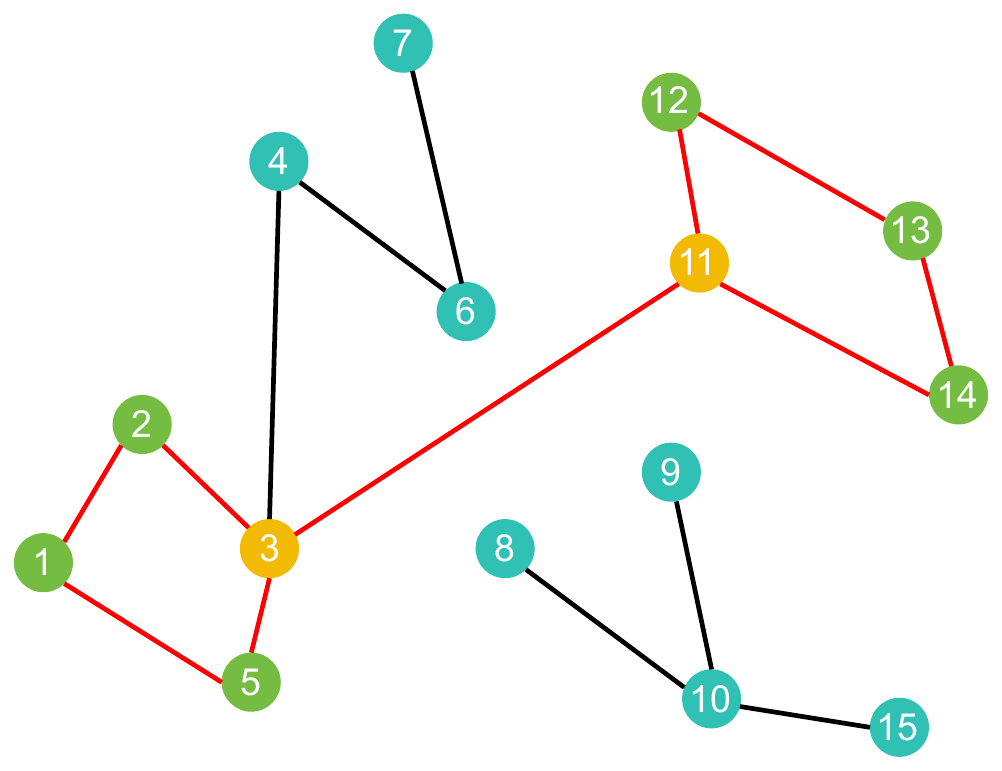}
\label{fig:15(d)}
\end{minipage}
}%
\subfigure[]{
\begin{minipage}[t]{0.3\linewidth}
\centering
\includegraphics[width=1\textwidth]{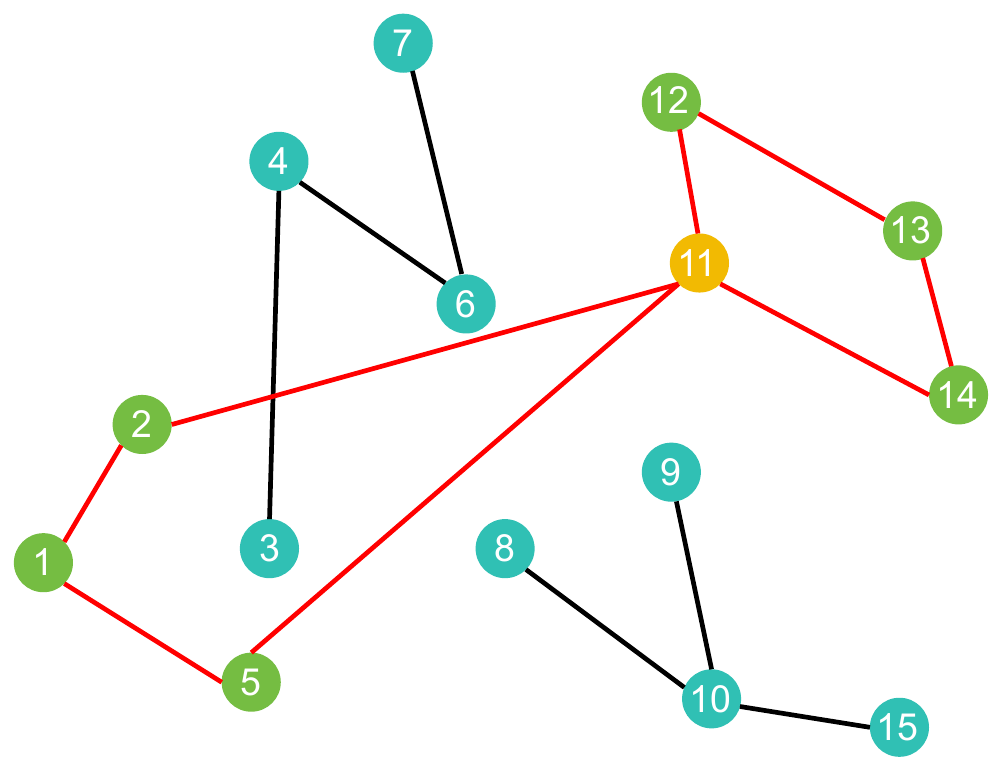}
\label{fig:15(e)}
\end{minipage}
}%
\subfigure[]{
\begin{minipage}[t]{0.3\linewidth}
\centering
\includegraphics[width=1\textwidth]{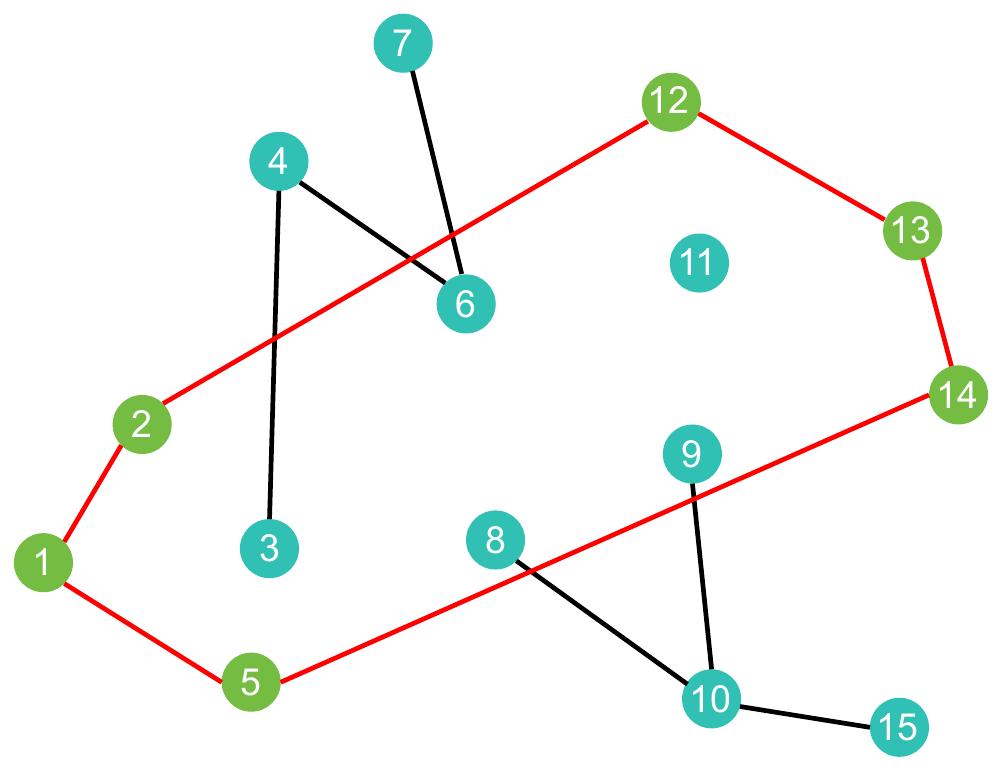}
\label{fig:15(f)}
\end{minipage}%
}%
\centering
\caption{The schematic diagram of distributing a GHZ state among the selected nodes marked in green. (a) The Steiner tree connecting them in the network with red edges. (b)-(f) According to the process of distributing entangled states in the tree network, the corresponding quantum walk and measurement operations are carried out at the corresponding nodes. Finally distribute a GHZ state among nodes 1, 2, 5, 12, 13 and 14.}
\label{fig:15}
\end{figure}

Finally, we provide an example of a concrete quantum network with $14$ nodes and describe the process of distributing GHZ states to several selected nodes, as shown in the \Cref{fig:15}. Select nodes 1, 2, 5, 12, 13 and 14 and mark them in green in the network diagram. Then, find the Steiner tree connecting them in the network and mark the additional points in yellow and the edge of the Steiner tree in red. The remaining points are marked in blue in \Cref{fig:15(a)}. For the Steiner tree, we select node 8 as the root node and perform the corresponding quantum walk operations at nodes 6 and 9. Then, nodes 8 and 3 and nodes 8 and 9 share entangled states as shown in \Cref{fig:15(b)}. We then perform the corresponding quantum walk operation on node 8, so that nodes 3 and 11 share an entangled state as shown in \Cref{fig:15(c)}. Then, we perform the corresponding quantum walk operations on nodes 3 and 11, and thus distribute two GHZ states nodes 1, 2, 3 and 5, and nodes 11, 12, 13 and 14, respectively, as shown in \Cref{fig:15(d)}. The corresponding quantum walk operation on node 3 is then performed, thus, nodes 1, 2, 5 and 11 share a GHZ state, as shown in \Cref{fig:15(e)}. Finally, a GHZ state is distributed among nodes 1, 2, 5, 12, 13 and 14, as shown in \Cref{fig:15(f)}, after performing the corresponding quantum walk operation on node 11.
Compared with \cite{59}, the quantum state that can distribute in an arbitrary quantum network is not only a 2-dimensional quantum state but also a $d$-dimensional GHZ state, which provides a more theoretical basis and experimental potential for high-dimensional quantum communication networks. 

\section{Application for constructing quantum fractal networks}

\begin{figure}[htbp]
\centering
\subfigure[G(0)]{
\begin{minipage}[t]{0.1\linewidth}
\centering
\includegraphics[width=0.9\textwidth]{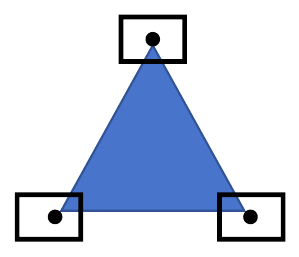}
\label{fig:8(a)}
\end{minipage}%
}%
\subfigure[G(1)]{
\begin{minipage}[t]{0.2\linewidth}
\centering
\includegraphics[width=0.9\textwidth]{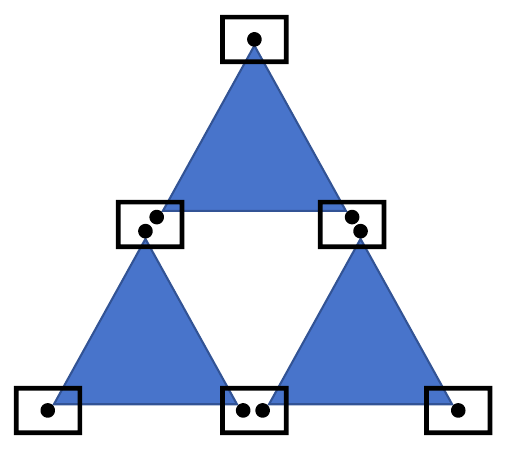}
\label{fig:8(b)}
\end{minipage}%
}%
\subfigure[G(2)]{
\begin{minipage}[t]{0.3\linewidth}
\centering
\includegraphics[width=0.9\textwidth]{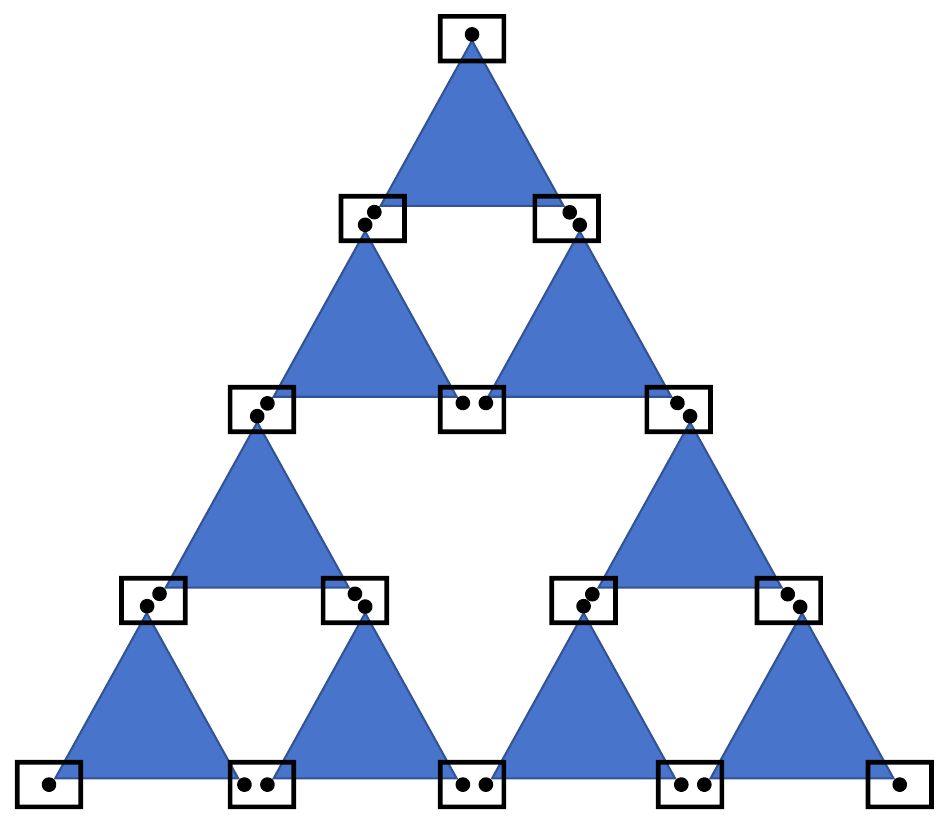}
\label{fig:8(c)}
\end{minipage}%
}%

\subfigure[]{
\begin{minipage}[t]{0.3\linewidth}
\centering
\includegraphics[width=0.9\textwidth]{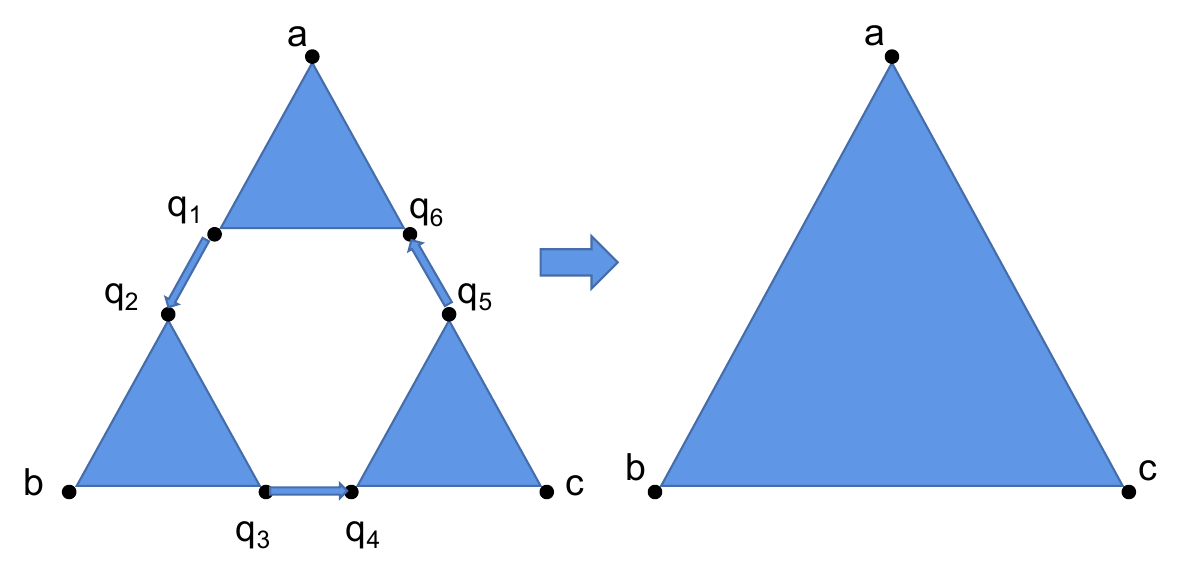}
\label{fig:9}
\end{minipage}%
}
\subfigure[]{
\begin{minipage}[t]{0.6\linewidth}
\centering
\includegraphics[width=0.9\textwidth]{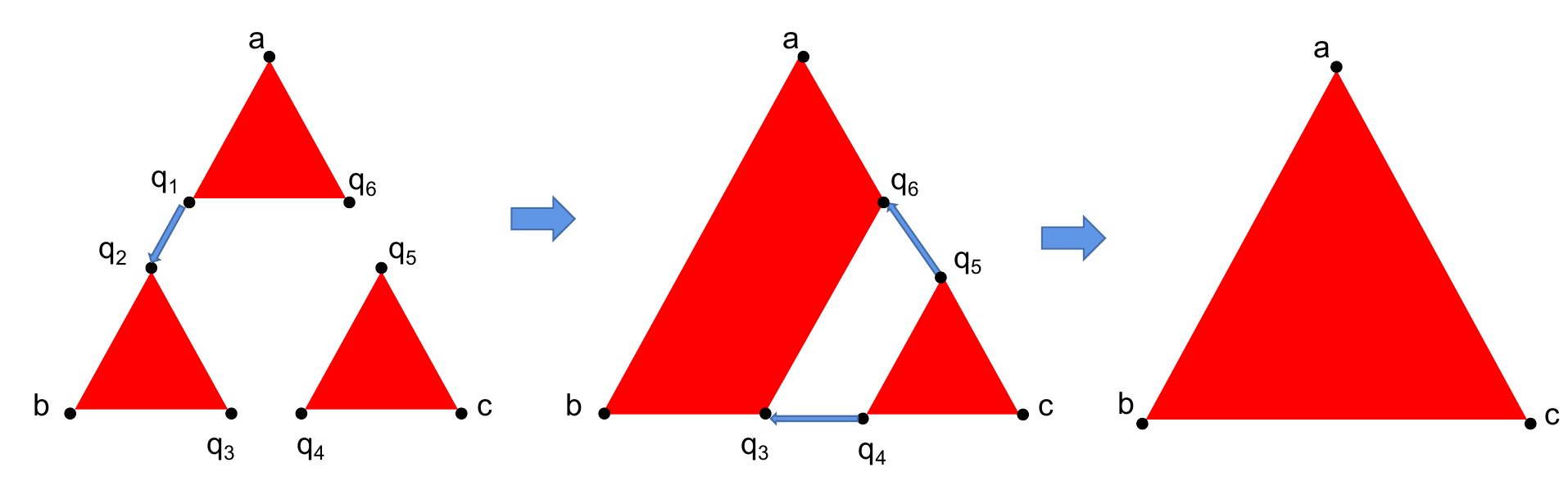}
\label{fig:10}
\end{minipage}%
}%

\subfigure[F(0)]{
\begin{minipage}[t]{0.1\linewidth}
\centering
\includegraphics[width=0.9\textwidth]{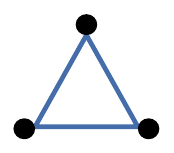}
\label{fig:11(a)}
\end{minipage}%
}%
\subfigure[F(1)]{
\begin{minipage}[t]{0.2\linewidth}
\centering
\includegraphics[width=0.9\textwidth]{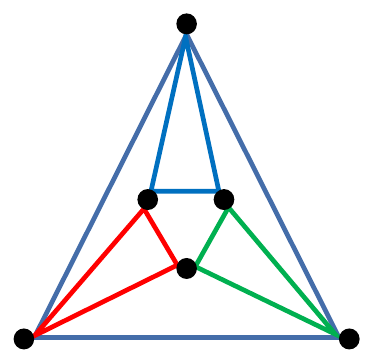}
\label{fig:11(b)}
\end{minipage}%
}%
\subfigure[F(2)]{
\begin{minipage}[t]{0.3\linewidth}
\centering
\includegraphics[width=0.9\textwidth]{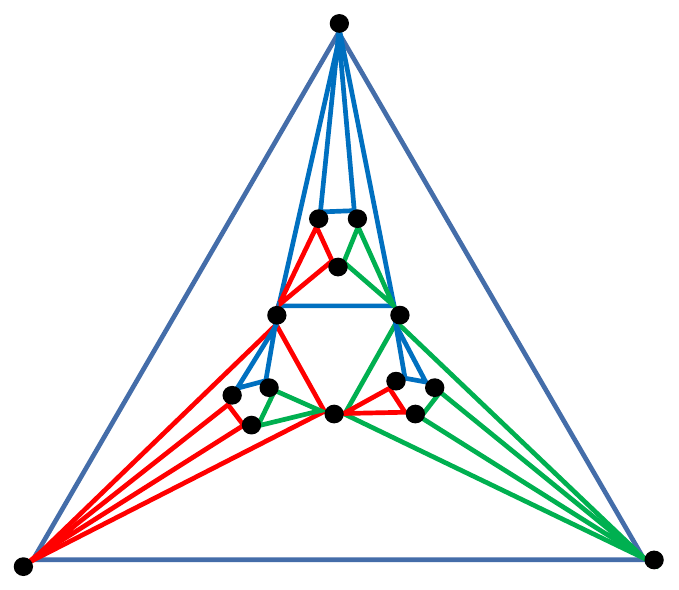}
\label{fig:11(c)}
\end{minipage}%
}%
\centering
\caption{(a)-(c) Some Sierpinski gaskets. Each black rectangle represents a quantum node. Each shaded triangle corresponds to a $3$-particle GHZ state between three nodes. (d) The entanglement distribution of a GHZ state between nodes $a, b, c$ corresponding to $G(1)$. Every blue shade triangle corresponds to a $3$-qubit GHZ state. (e) The entanglement distribution of a $3$-qudit GHZ states corresponding to $G(1)$. Every red shade triangle corresponds to a $3$-qudit GHZ state. (f)-(h) Some evolutionary fractal networks constructed from the Sierpinski gaskets. Black dots represent network nodes. Each triangle represents a three-party channel based on a GHZ entangled state. The triangles in the figure are colored red, green and blue to distinguish different tripartite channels based on different GHZ States.}
\end{figure}

We constructed a quantum fractal network using the aforementioned entanglement distribution scheme. Inspired by the approach used to create classical complex networks from fractal structures \cite{48,49}, we can extend this methodology to construct a quantum network utilizing the Sierpinski gasket, as shown in \Cref{fig:8(a),fig:8(b),fig:8(c)}. Each black rectangle represents a quantum node. Each shaded triangle represents a 3-particle GHZ state, and the three particles of each GHZ state are located at three quantum nodes, forming the structure of the Sierpinski gasket. Let the side length of each small equilateral triangle be $L$, the side length of $G(N)$, the Sierpinski gasket with N iterations, is $2^N L$.

Unlike the method provided by \cite{60,61} for using Bell state measurements to connect three GHZ states, we can distribute a $3$-qubit GHZ state corresponding to a shaded triangle with side length $2L$ using three $3$-qubit GHZ states corresponding to three shaded triangles with side length $L$ through the operations in \Cref{fig:9}, where an arrow represents a quantum walk operation, and the starting and target points of the arrows represent coin states and position states, respectively.
We let the initial state in \Cref{fig:9} be
\begin{align}
|\Psi(0)\rangle=\frac{|000\rangle+|111\rangle}{\sqrt{2}}_{a,q_1,q_6}\otimes\frac{|000\rangle+|111\rangle}{\sqrt{2}}_{q_2,b,q_3}\otimes\frac{|000\rangle+|111\rangle}{\sqrt{2}}_{q_4,q_5,c}.
\end{align}
After the quantum walks in \Cref{fig:9}, $C=X$, and $S$ is the CNOT gate. The quantum state becomes
\begin{align}
&|\Psi(1)\rangle=\frac{1}{2\sqrt{2}}(|111\rangle(|111\rangle|000\rangle+|000\rangle|111\rangle)+|010\rangle(|110\rangle|001\rangle+|001\rangle|110\rangle)\nonumber\\
&+|100\rangle(|101\rangle|010\rangle+|010\rangle|101\rangle)+|001\rangle(|011\rangle|100\rangle+|100\rangle|011\rangle))_{q_6,q_2,q_4,q_1,q_3,q_5,a,b,c}.
\end{align}
Then, we take a measurement with the basis $\{|+\rangle, |-\rangle\}$ in the particle $q_1, q_3, q_5$ and the basis $\{|0\rangle, |1\rangle\}$ in the particles $q_2, q_4, q_6$. The entanglement states of the particles $a, b, c$ can be recovered to the state $\frac{|000\rangle+|111\rangle}{\sqrt{2}}$ by some local unitary operations according to the measurement results.

Moreover, our method is not limited to a $2$-dimensional case like the method in \cite{60,61}. Thus, we can generalize our method to a d-dimensional case. This not only avoids the experimental technical difficulty of high-dimensional joint Bell state measurement but also greatly improves the communication efficiency of quantum networks by using high-dimensional GHZ states. The initial state in \Cref{fig:10} is: 
\begin{align}
|\Psi(0)\rangle=\frac{1}{\sqrt{d}}\sum\limits_{i_1=0}^{d-1}|i_1, i_1, i_1\rangle_{a,q_1,q_6}\otimes\frac{1}{\sqrt{d}}\sum\limits_{i_2=0}^{d-1}|i_2, i_2, i_2\rangle_{q_2,b,q_3}\otimes\frac{1}{\sqrt{d}}\sum\limits_{i_3=0}^{d-1}|i_3, i_3, i_3\rangle_{q_4,q_5,c}.
\end{align}
After the first quantum walk in the \Cref{fig:10} with the coin operator $C=I$, and the conditional shift operator $S=\sum\limits_{k,j=0}^{d-1}|k\rangle\langle k|\otimes|j-k\rangle\langle j|$. Then, we performed the measurement with the basis $\{|\widetilde{k}\rangle: k=0,\cdots,d-1\}$ and $\{|k\rangle: k=0,\cdots,d-1\}$ on the particles $q_1$ and $q_2$, respectively. If the measurement results of the particles $q_1$ is $\widetilde{p_1}$ and the measurement results of the particles $q_2$ is $u_1$, then the entanglement state of leftover particles is 
\begin{align}
|\Psi(1)\rangle=\frac{1}{\sqrt{d}}\sum\limits_{i_1=0}^{d-1}\omega^{-i_1\widetilde{p_1}}|i_1,i_1,i_1+u_1,i_1+u_1\rangle_{a,q_6,b,q_3}\otimes\frac{1}{\sqrt{d}}\sum\limits_{i_3=0}^{d-1}|i_3, i_3, i_3\rangle_{q_4,q_5,c}.
\end{align}
After the last two quantum walks in the \Cref{fig:10} with the coin operator $C=I$, and the conditional shift operator $S=\sum\limits_{k,j=0}^{d-1}|k\rangle\langle k|\otimes|j-k\rangle\langle j|$. Then, we take a measurement with the basis $\{|\widetilde{k}\rangle: k=0,\cdots,d-1\}$ and $\{|k\rangle: k=0,\cdots,d-1\}$ on particles $q_4,q_5$ and $q_3, q_6$, respectively. If the measurement results of particles $q_4,q_5$ are $\widetilde{p_2},\widetilde{p_3}$. The measurement results for particles $q_6, q_3$ are $u_2, u_3$, and $u_1+u_2=u_3$. The entanglement state of the remaining particles is: 
\begin{eqnarray}
|\Psi(2)\rangle=\sum\limits_{i_1=0}^{d-1}\frac{\omega^{-i_1(\widetilde{p_1}+\widetilde{p_2}+\widetilde{p_3})+u_2(\widetilde{p_2}+\widetilde{p_3})}}{\sqrt{d}}|i_1,i_1+u_1,i_1-u_2\rangle_{a,b,c}.
\end{eqnarray}
The entanglement states of particles $a, b, c$ corresponding to different measurement results can be recovered to the state $\frac{1}{\sqrt{d}}\sum\limits_{i_1=0}^{d-1}|i_1, i_1, i_1\rangle$ by some local unitary operations according to the measurement results.

Furthermore, we can distribute a $3$-qubit or $3$-qudit GHZ state corresponding to a shaded triangle with side length $2^N L$ using the GHZ states corresponding to the shaded triangles in $G(N)$ through $\frac{3^N-1}{2}$ operations in \Cref{fig:9} or \Cref{fig:10}. Therefore, we can construct a complex quantum network between quantum nodes by distributing entanglement between these quantum nodes. Inspired by classical complex networks in real life that generally have self-similarity\cite{50}, we can construct a quantum complex network according to the structure of the Sierpinski gasket, which will bring some self-similarity to the actual quantum network structure. 

\begin{figure}[htbp]
\centering
\subfigure[]{
\begin{minipage}[t]{0.33\linewidth}
\centering
\includegraphics[width=1\textwidth]{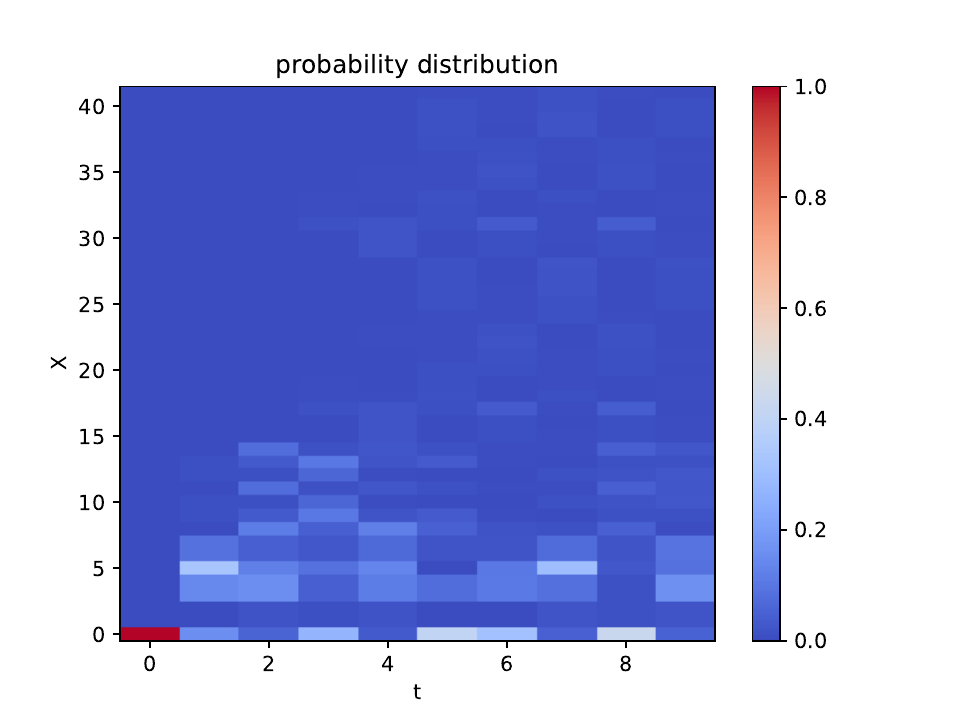}
\label{fig:s2}
\end{minipage}%
}%
\subfigure[]{
\begin{minipage}[t]{0.33\linewidth}
\centering
\includegraphics[width=1\textwidth]{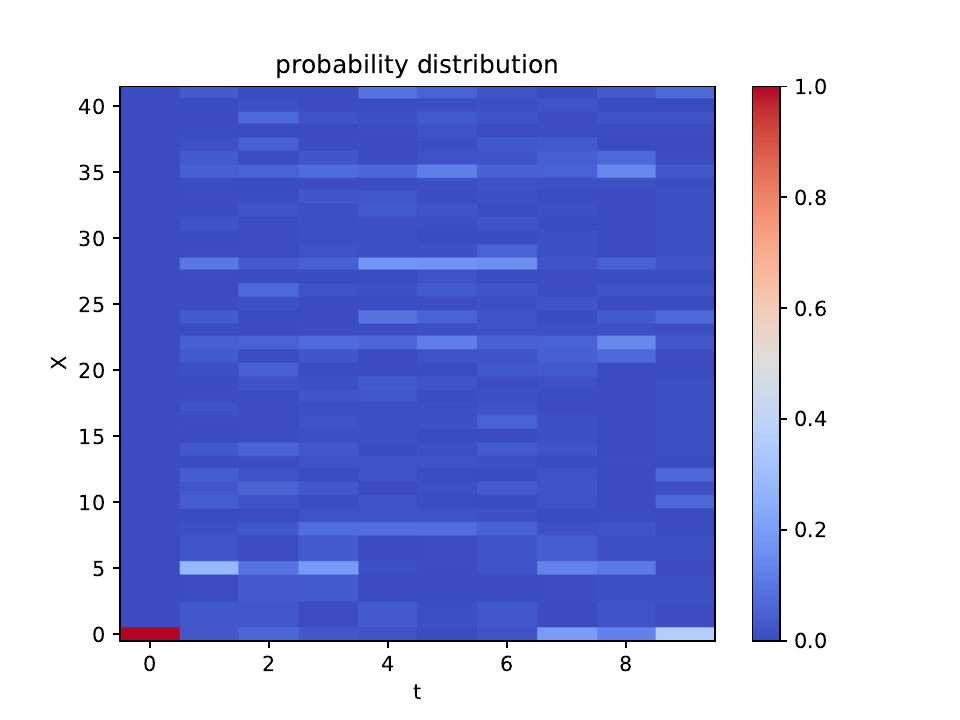}
\label{fig:s3}
\end{minipage}%
}%
\subfigure[]{
\begin{minipage}[t]{0.33\linewidth}
\centering
\includegraphics[width=1\textwidth]{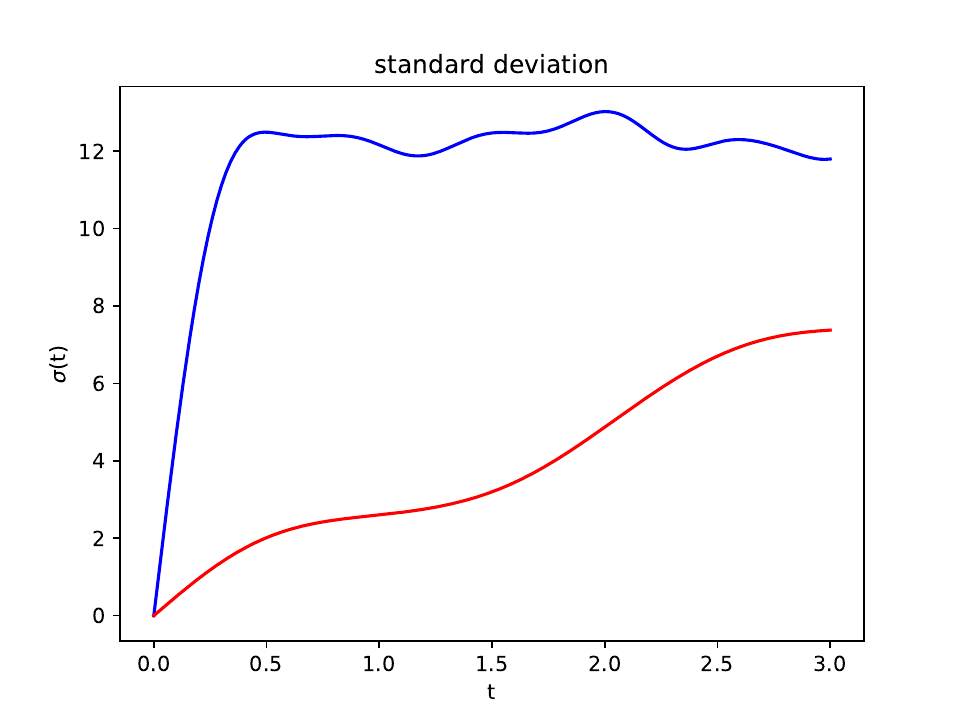}
\label{fig:s4}
\end{minipage}
}%

\centering

\caption{(a)-(b) The probability distribution of the particle being at each node after continuous quantum walks of time $t$ over the two graph structures, starting from the topmost node. (a) The Sierpinski gasket. (b) Our quantum network. (c) The standard deviation function of the continuous quantum walks of time $t$. Red and blue lines responds to the Sierpinski gasket and our quantum network,respectively.}
\label{fig:s}
\end{figure}

From $G(N)$, we can construct the corresponding quantum network $F(N)$. The construction rule is as follows: the GHZ entangled states corresponding to the basic shaded triangles in $G(N)$ are regarded as basic entangled states, and we assume that these basic entangled states can be prepared directly and continuously. The three nodes, where the particles of the basic entangled state are located, are entangled by the basic entangled state. Using these basic entangled states, we can prepare GHZ states with particles farther apart so that some nodes that are not originally entangled by the same entangled state can be entangled by the same entangled state. In the constructed network, we can establish a tripartite channel between three nodes that can be entangled by the same GHZ entangled state, as shown in \Cref{fig:11(a),fig:11(b),fig:11(c)}. 
Research on complex quantum networks has attracted considerable attention\cite{75}. On the one hand, complex quantum networks leverage quantum algorithms to demonstrate significant attributes of complex networks. For example, continuous quantum walks in fractal networks exhibit a series of quantum transport properties related to fractal networks\cite{76,77,78}. We studied the quantum transport properties of continuous quantum walks in newly designed quantum fractal networks. Compared with direct continuous quantum walks on Sierpinski gaskets, the quantum walks on the new network diffuse faster within the same period of time, which can be clearly seen from the probability distribution and standard deviation of the positions, as shown in \Cref{fig:s2,fig:s3,fig:s4}. Because we increased the number of edges between nodes according to the network design method, this led to an acceleration of quantum diffusion in this network compared to the Sierpinski gasket.

In addition, the characteristics of complex networks have been leveraged to advance the development of quantum algorithms and quantum technologies. The GHZ state plays a vital role in many quantum information science applications such as quantum teleportation, quantum signature schemes\cite{72,73}, and quantum secure communication\cite{74}. Our designed quantum fractal network, based on the fractal structure of the Sierpinski gasket, can effectively distribute GHZ states among nodes. Moreover, based on the GHZ states among nodes and existing schemes, we can implement quantum information technologies such as quantum teleportation and quantum secure direct communication on this network.
As for the constructed quantum network, we describe its classical complex network properties in more detail in the appendix A.

\section{Experiments}

\begin{figure*}[htbp]
\centering
\subfigure[]{
\begin{minipage}[t]{0.4\linewidth}
\centering
\includegraphics[width=1\textwidth]{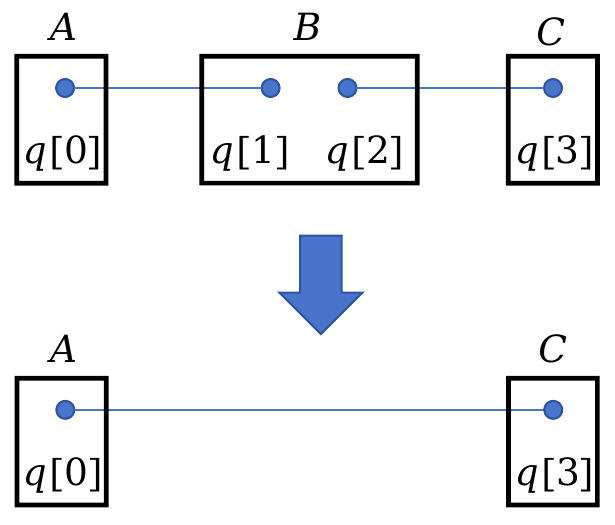}
\label{fig:25}
\end{minipage}%
}%
\subfigure[]{
\begin{minipage}[t]{0.4\linewidth}
\centering
\includegraphics[width=1\textwidth]{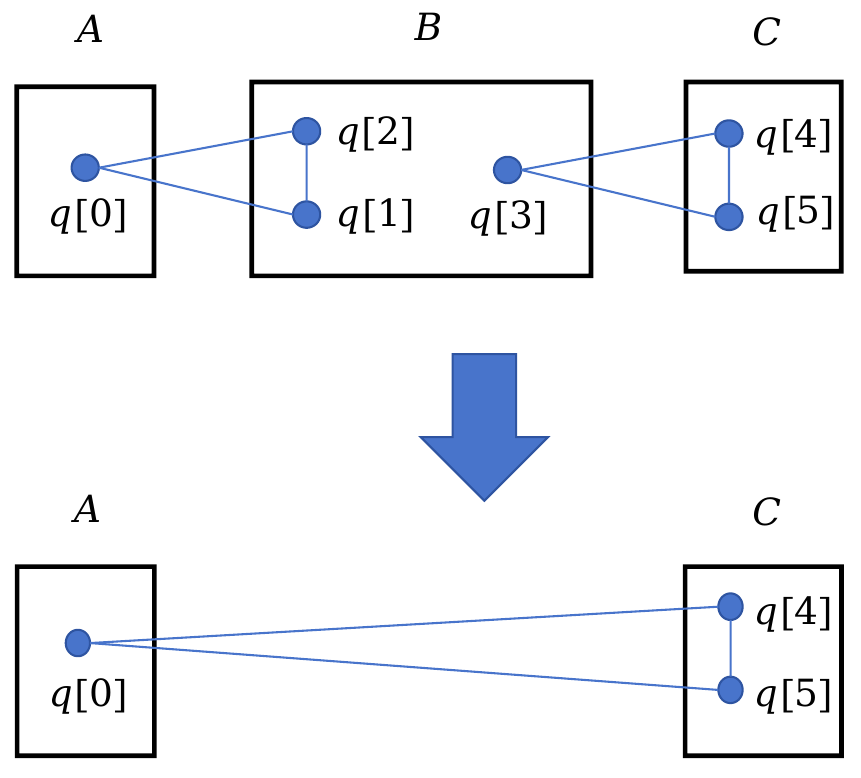}
\label{fig:26}
\end{minipage}%
}%

\subfigure[]{
\begin{minipage}[t]{0.45\linewidth}
\centering
\includegraphics[width=1\textwidth]{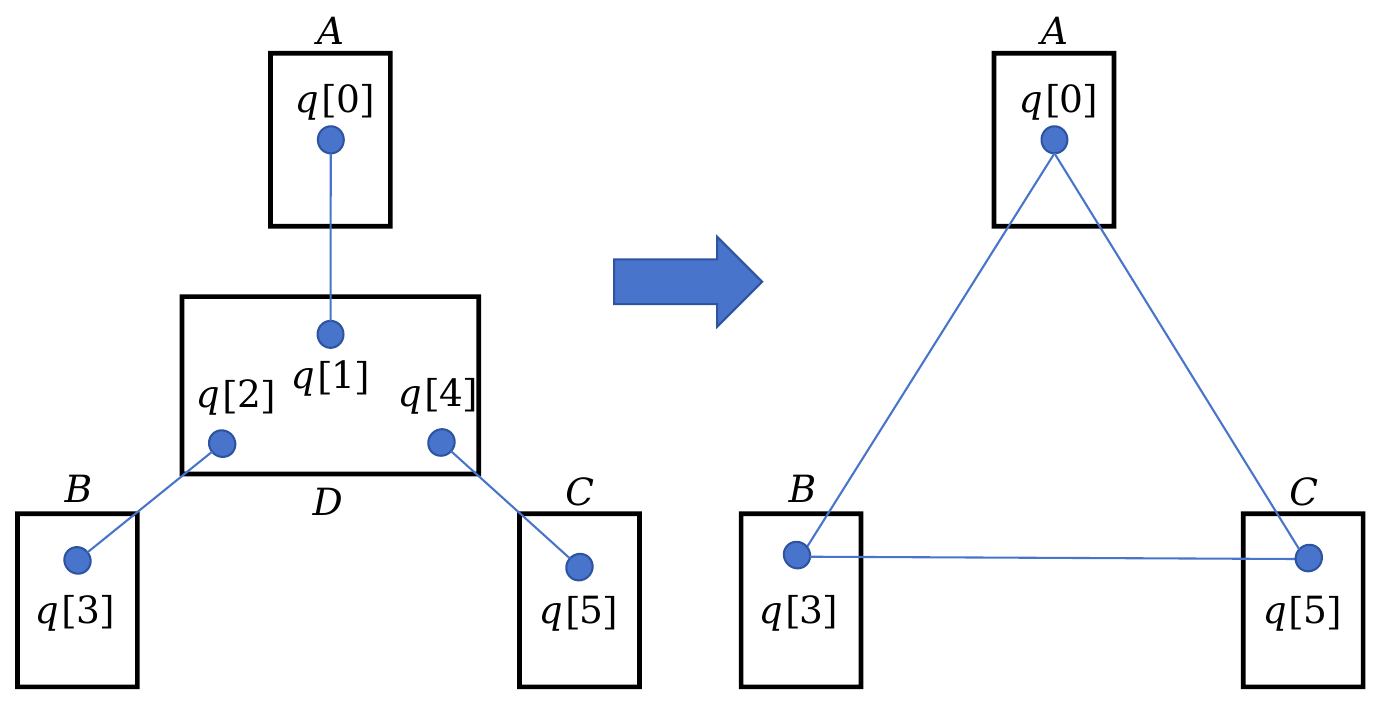}
\label{fig:27}
\end{minipage}%
}%
\subfigure[]{
\begin{minipage}[t]{0.45\linewidth}
\centering
\includegraphics[width=1\textwidth]{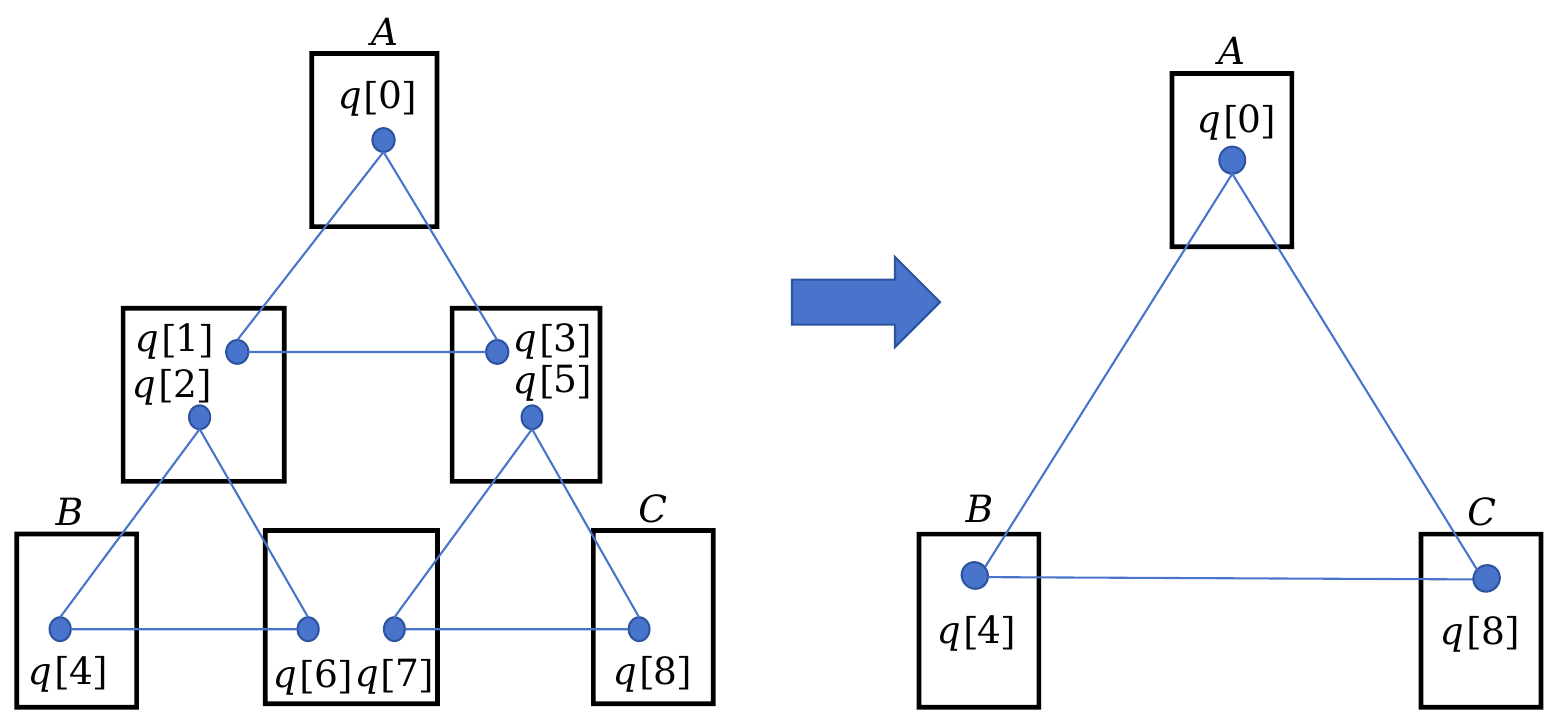}
\label{fig:28}
\end{minipage}%
}%

\centering

\caption{The schematic diagram of some modules implemented in experiments. (a)-(b) The schematic diagram of 2-party entanglement distributions using two Bell states and GHZ states, respectively. (c)-(d) The schematic diagram of 3-party entanglement distributions using three Bell states and GHZ states, respectively.}
\label{fig:2528}
\end{figure*}

\begin{figure*}[ht]
\centering
\includegraphics[width=0.85\textwidth]{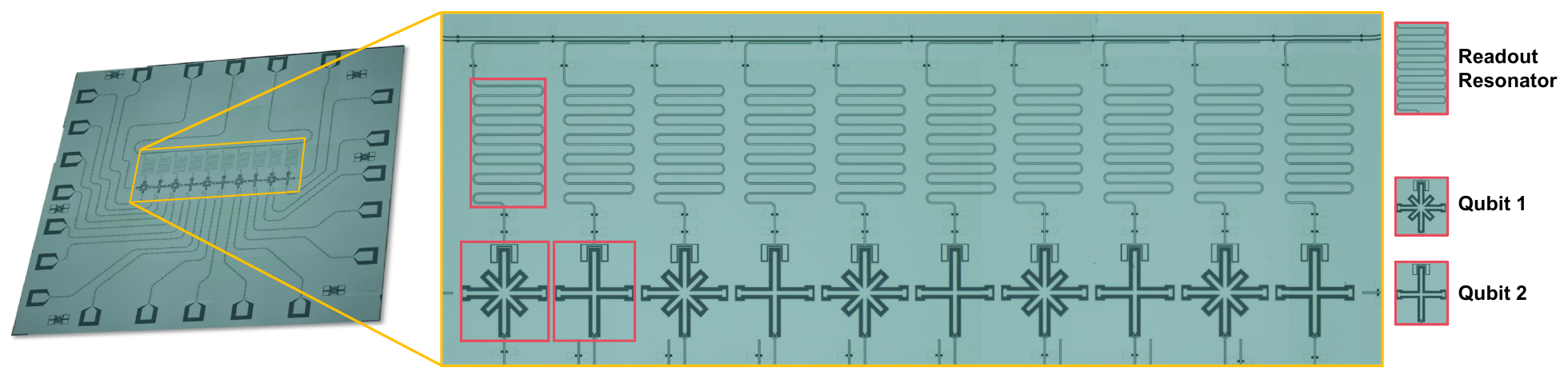}
\caption{The schematic diagram of the quantum superconducting processor. Ten transmon qubits are integrated on the quantum processor as a one-dimensional chain, and adjacent qubits are directly coupled through capacitance. It shows two types of qubits with different shapes of capacitors, resulting in different sweet spots of approximately 500MHz. In our experiments, each qubit exists as an artificial two-level system.}
\label{fig:16}
\end{figure*}

We implemented some experiments of some modules (\Cref{fig:2528}) of entanglement distributions on a superconducting quantum processor, which shows the feasibility of our theory in physical experiments to some extent. A schematic of the quantum processor is shown in \Cref{fig:16}. In these studies, the circuits we used consist of 10 qubits in a one-dimensional array. The Hamiltonian of this system is written as follows with the rotating wave approximation:
\begin{eqnarray}
H/\hbar =\sum_{i}  -\frac{1}{2}\omega_{i}\sigma^{z}_{i} + 
 \sum_{i<j} g_{ij}(\sigma^{+}_{i}\sigma^{-}_{i+1}+h.c.)
\end{eqnarray}
where $\omega_i$ is the frequency of qubit $Q_i$, $g_{ij}$ is the coupling coefficient of qubits $Q_i,Q_j$, and $\sigma^{\pm}$ is the raising (lowering) operator.

Only the nearest-neighbor (NN) qubits are directly coupled through capacitance, with a coupling strength range from 10 to 12 MHz, which is at least one order of magnitude larger than that of non-NN qubits. The two-qubit control-Z gates of NN qubits are achieved within 50 ns with the non-adiabatic scheme adopted, while those of non-NN qubits cannot be achieved due to the coupling strength. The frequency of each qubit can be tuned-up from 4.0 GHz to 5.6 GHz through a microwave line, and excitation can be realized by a microwave drive line for each qubit. All qubits are coupled to one readout line through their individual readout resonators, the frequencies of which are from 6.49 to 6.66 GHz. Readout is achieved through one shared readout line coupled to readout resonators of all qubits. Other information of the superconducting quantum processor is provided in Appendices B-D. For each experiment, we show the result with the highest average fidelity among the five repetitions by plotting the density matrix obtained from the experiment in the Supplementary Information.

\subsection{Two-party entanglement distributions}
\subsubsection{Using two Bell states}

We distribute a Bell state between $A$ and $B$ nodes using two Bell states as shown in \Cref{fig:25}. We used a digital quantum circuit to demonstrate the entire experimental process. The entire experiment was divided into three stages: preparation, quantum walk and measurement. In the preparation stage, we prepared two Bell states between $q[0]$ and $q[1]$, $q[2]$ and $q[3]$, respectively. In the quantum walk stage, we performed the corresponding quantum walk between $q[1]$ and $q[2]$ according to the basic modules. In the measurement stage, we measured $q[1]$ and $q[2]$ in proper basis and obtain a new Bell state. In the \Cref{fig:17(a)}, the quantum circuits were recompiled into pulse form, and the pulse sequences were sensed by real qubits corresponding to the quantum circuit. On considering the way qubits connected due to the geometric structure, a two-qubit Control-Z gate can only be formed between adjacent qubits. In our all experiments, the rotation around the z-axis was achieved by virtual Z which has been omitted in the schematic diagrams of pulse sequences for clarity, and single qubit rotation was achieved using microwave pulses with the same frequency as the idle point of the qubit. Two approximately square wave pulses were used to achieve CZ gates between adjacent qubits. The orange line represents the pulse acting on the coin qubit. The gray line represents the pulse acting on the position qubit. The blue line represents the pulse acting on the desired qubits. In this work, $q[0], q[1], q[2], q[3]$ in quantum circuits were mapped to qubits $Q_4, Q_5, Q_7, Q_6$ on the real quantum processor, respectively. We finally obtained the fidelity of the new entangled state between $q[0]$ and $q[3]$ through quantum state tomography\cite{47} corresponding to measurement results of $q[1]$ and $q[2]$. We ran this experiment with 10000 shots on the superconducting quantum processor. The results are shown in the \Cref{fig:17(b)}. Through these fidelity, we can get average fidelity $F=79.8(2.0)\%$

\subsubsection{Using two GHZ states}
We distributed a GHZ state between $A$ and $B$ nodes using two GHZ states, as shown in \Cref{fig:26}. For the distribution of a 2-dimensional GHZ state, we have two methods. One is to directly use the 2-dimensional GHZ state module in \Cref{fig:6(a)}, and the other is to use the d-dimensional GHZ state module in \Cref{fig:7(b)} to set $d=2$. 

\textbf{method 1} According to the 2-dimensional GHZ state entanglement distribution module in \Cref{fig:6(a)}, we constructed a quantum circuit and recompiled it into pulse form, as shown in \Cref{fig:18(a)}, where $q[0], q[1], q[2], q[3], q[4], q[5]$ in quantum circuits were mapped to qubits $Q_4, Q_5, Q_7, Q_6, Q_8, Q_9$ on the real quantum processor. In the preparation stage, we prepared two GHZ states between $q[0], q[1], q[2]$ and $q[3], q[4], q[5]$, respectively. In the quantum walk stage, we performed the corresponding quantum walk between $q[1], q[2]$ and $q[3]$ according to the module. In the measurement stage, we measured $q[1], q[2]$ and $q[3]$ in proper basis and distributed a new GHZ state between $q[0], q[4]$ and $q[5]$.  We ran the quantum circuit with 10000 shots on the superconducting quantum processor. Through the quantum state tomography, we can obtain the fidelity of the entangled state among $q[0], q[4]$ and $q[5]$ corresponding to any measurement result of $q[1], q[2]$ and $q[3]$. The fidelity results are shown in \Cref{fig:18(b)}. Through these fidelity, we can get average fidelity $F=54.3(3.6)\%$

\textbf{method 2} According to the above $d$-dimensional GHZ state entanglement distribution module in \Cref{fig:7(b)}, we can construct a quantum circuit and recompile it into pulse form, as shown in \Cref{fig:19(a)}, where $q[0], q[1], q[2], q[3], q[4], q[5]$ in quantum circuits were mapped to qubits $Q_4, Q_5, Q_7, Q_6, Q_8, Q_9$ on the real quantum processor, respectively. In the preparation stage, we prepared two GHZ states between $q[0], q[1], q[2]$ and $q[3], q[4], q[5]$, respectively. In the quantum walk \& local recover operator stage, we performed the corresponding quantum walk between $q[1], q[2]$ and $q[3]$ according to the above theory and perform a local recover operator on $q[0]$. In the measurement stage, we measured $q[1], q[2]$ and $q[3]$ in proper basis to distribute a GHZ state between $q[0], q[4]$ and $q[5]$.  We ran the quantum circuit with 10000 shots on the superconducting quantum processor. We obtained the fidelity of the entangled state among $q[0], q[4]$ and $q[5]$ corresponding to any measurement result of $q[1], q[2]$ and $q[3]$. The fidelity results are shown in the \Cref{fig:19(b)}. Through these fidelity, we can get average fidelity $F=54.3(3.2)\%$

\subsection{Three-party entanglement distributions}

According to the above basic entanglement distribution modules in \Cref{fig:7(c)} and \Cref{fig:9}, we can distribute a $2$-dimensional GHZ state using three Bell states or GHZ states to $A, B$ and $C$ nodes as shown in \Cref{fig:27} and \Cref{fig:28}. 

\subsubsection{Using three Bell states} 
According to the module in \Cref{fig:7(c)}, we designed the quantum circuit of a $2$-dimensional GHZ state entanglement distribution using three Bell states and recompiled it into pulse form, as shown in \Cref{fig:20(a)}, where $q[0], q[1], q[2], q[3], q[4], q[5]$ in the quantum circuits were mapped to qubits $Q_5, Q_8, Q_7, Q_9, Q_6, Q_4$ on the real quantum processor. In the preparation stage, we prepared three Bell states between $q[0]$ and $q[1]$, $q[2]$ and $q[3]$, $q[4]$ and $q[5]$, respectively. In the quantum walk \& local recover operator stage, we performed the corresponding quantum walk between $q[1], q[2]$ and $q[4]$ according to the above theory and perform a local recover operator on $q[3]$. In the measurement stage, we measured $q[1], q[2]$ and $q[4]$ in a proper basis and obtained a GHZ state between $q[0], q[3]$ and $q[5]$. We ran the quantum circuit with 10000 shots on the superconducting quantum processor. We can obtain the fidelity of the entangled state among $q[0], q[3]$ and $q[5]$ corresponding to any measurement result of $q[1], q[2]$ and $q[4]$. The fidelity results are shown in the \Cref{fig:20(a)}. Through these fidelity, we can get average fidelity $F=54.4(1.1)\%$

\subsubsection{Using three GHZ states}

\begin{figure*}[htbp]
\centering
\subfigure[]{
\begin{minipage}[t]{0.3\linewidth}
\centering
\includegraphics[width=1\textwidth]{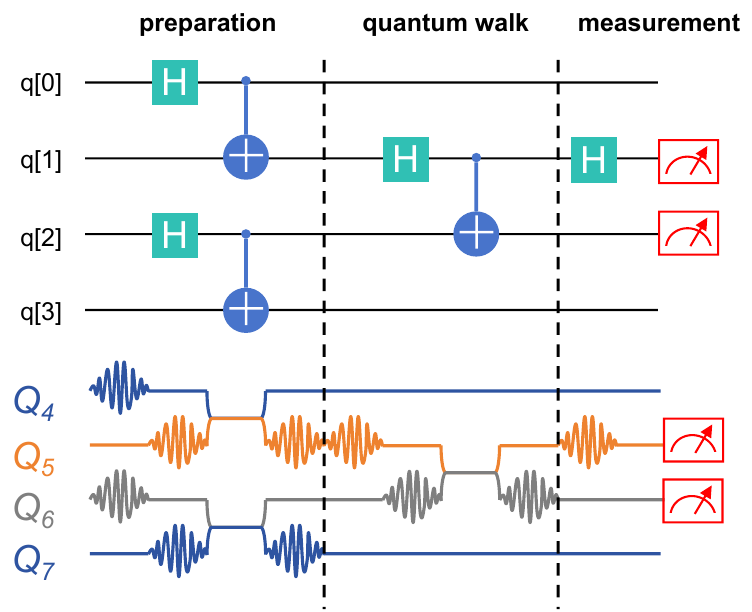}
\label{fig:17(a)}
\end{minipage}%
}%
\subfigure[]{
\begin{minipage}[t]{0.3\linewidth}
\centering
\includegraphics[width=1\textwidth]{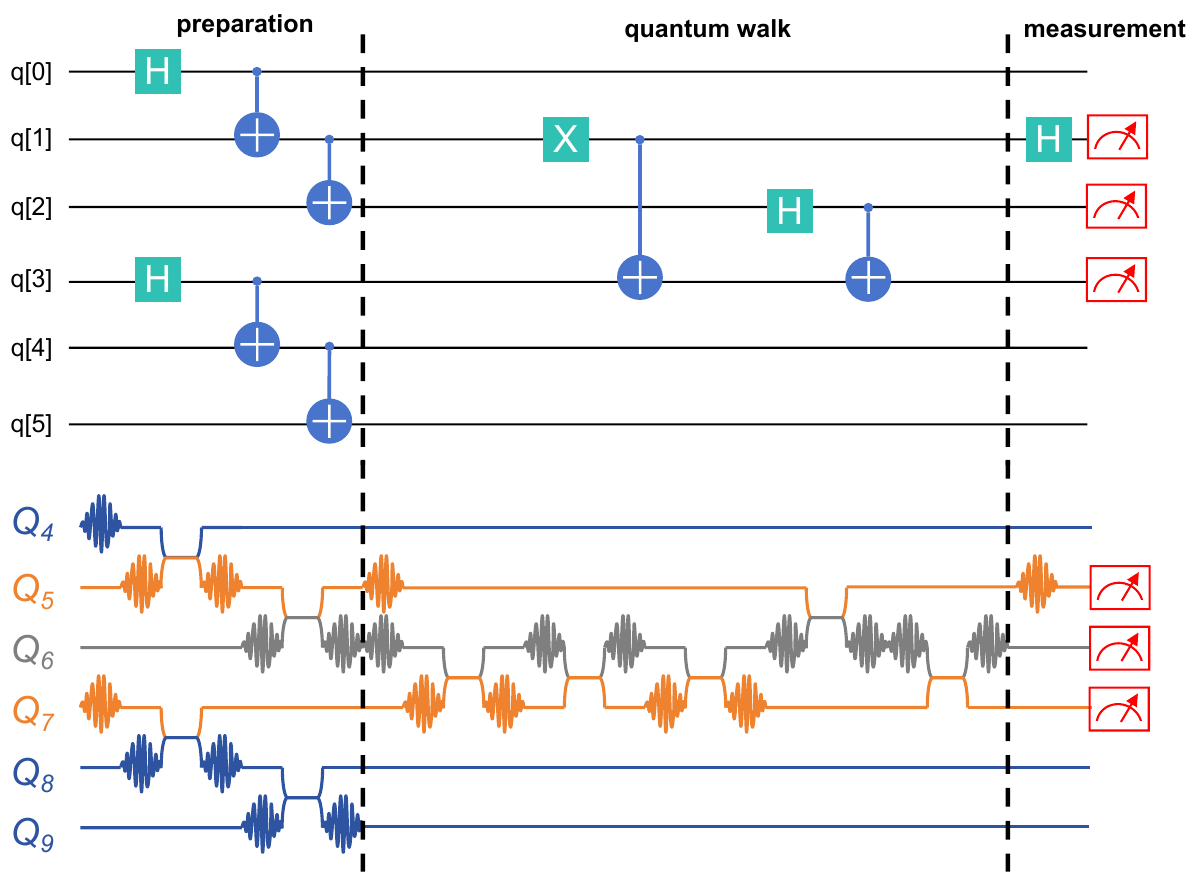}
\label{fig:18(a)}
\end{minipage}%
}%
\subfigure[]{
\begin{minipage}[t]{0.3\linewidth}
\centering
\includegraphics[width=1\textwidth]{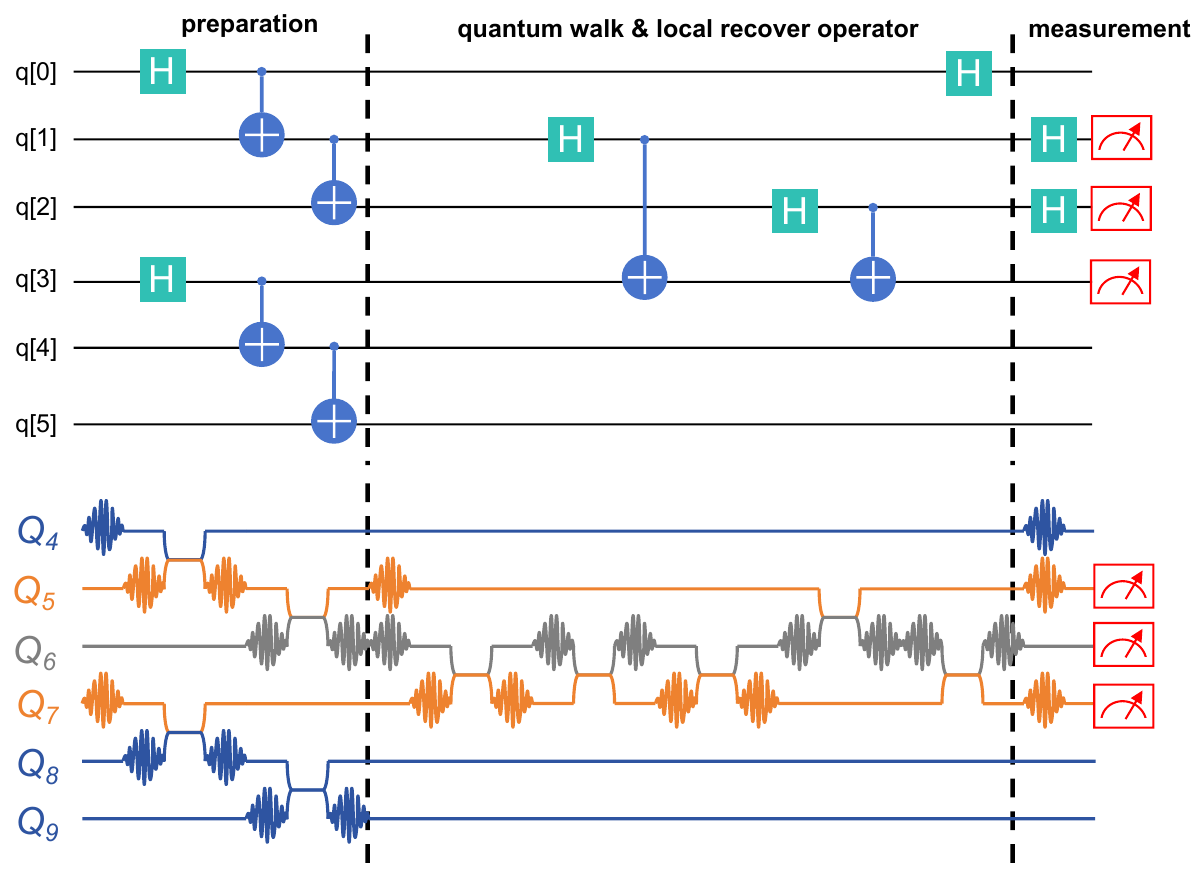}
\label{fig:19(a)}
\end{minipage}%
}%

\subfigure[]{
\begin{minipage}[t]{0.3\linewidth}
\centering
\includegraphics[width=1\textwidth]{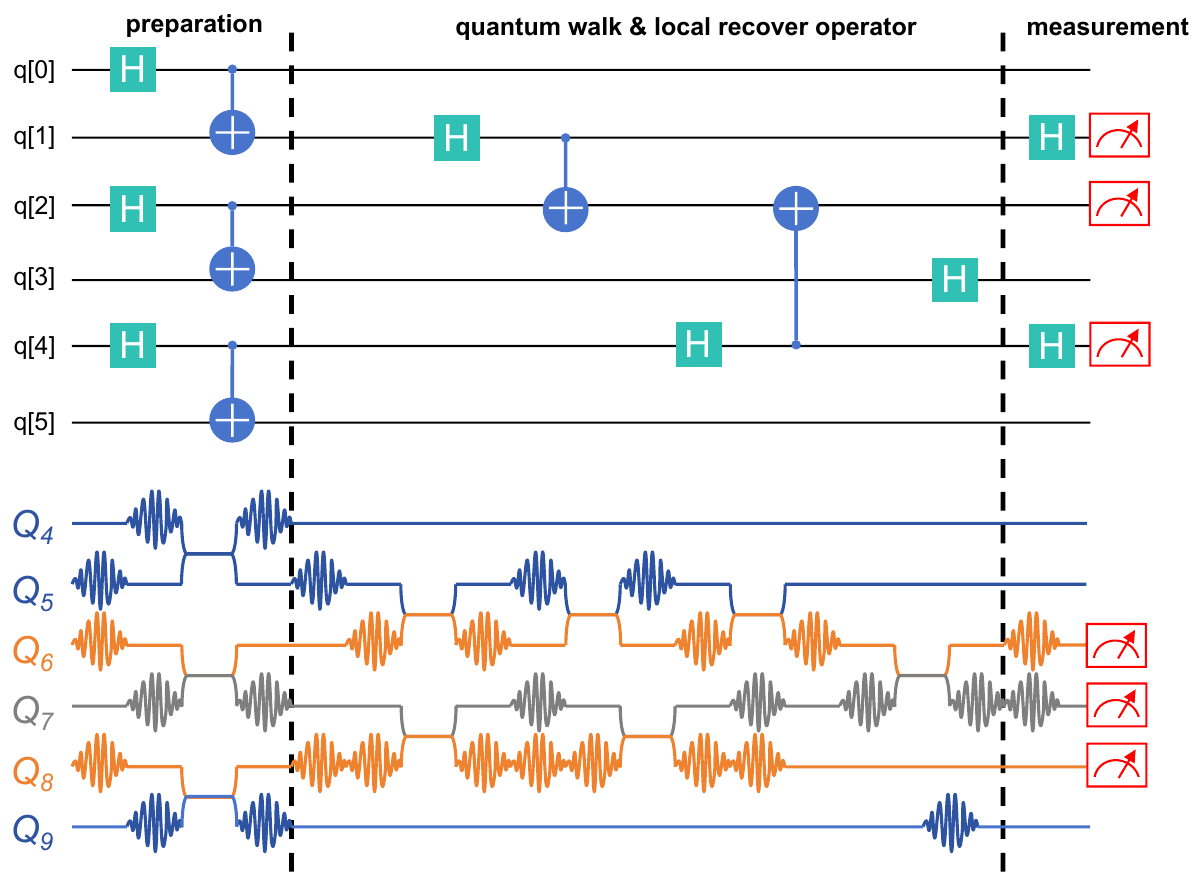}
\label{fig:20(a)}
\end{minipage}%
}%
\subfigure[]{
\begin{minipage}[t]{0.3\linewidth}
\centering
\includegraphics[width=1\textwidth]{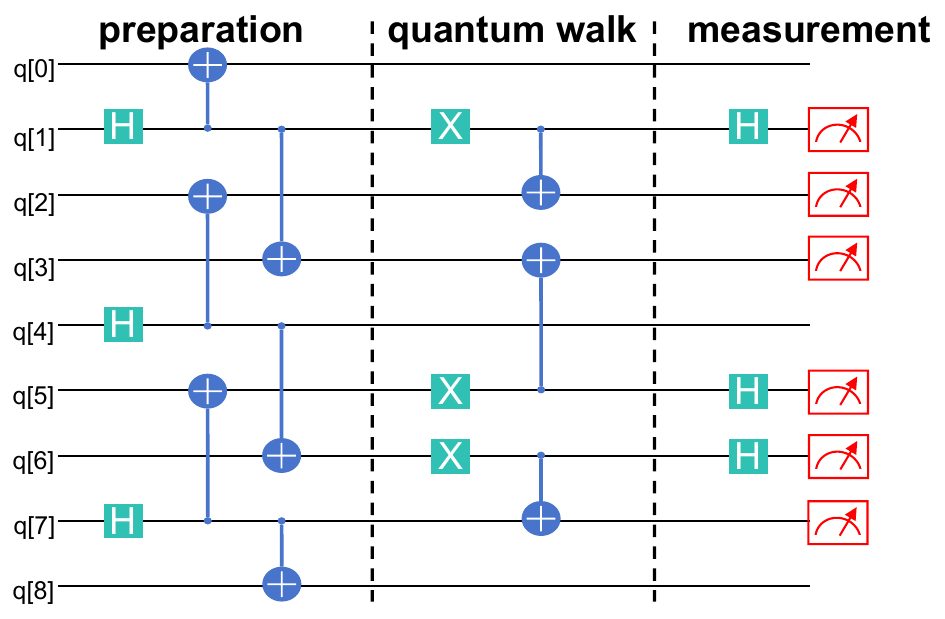}
\label{fig:21(a)}
\end{minipage}%
}%
\subfigure[]{
\begin{minipage}[t]{0.3\linewidth}
\centering
\includegraphics[width=1\textwidth]{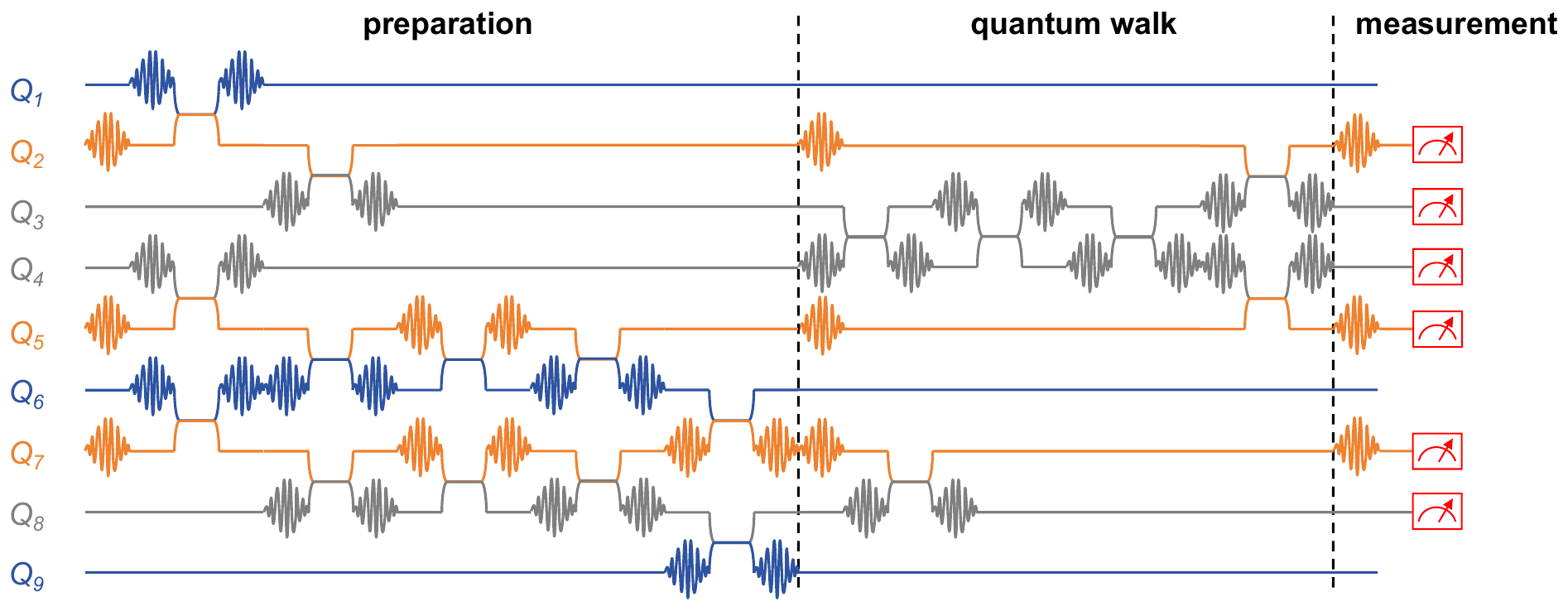}
\label{fig:21(b)}
\end{minipage}%
}%

\centering
\caption{(a) The schematic diagram of the quantum circuit and its pulse form of the entanglement swapping experiment for two 2-dimensional Bell states. (b) The schematic diagram of the quantum circuit and its pulse form of method 1 for the
entanglement swapping experiment of two 2-dimensional GHZ
states. (c) The schematic diagram of the quantum circuit and its pulse form of method 2 for the
entanglement swapping experiment of two 2-dimensional GHZ
states. (d) The schematic diagram of the quantum circuit and its pulse form of the entanglement
swapping experiment of three 2-dimensional Bell states. (e)-(f) The schematic diagram of the quantum circuit and its pulse form of the entanglement
swapping experiment of three 2-dimensional GHZ states.}
\end{figure*}

\begin{figure*}[htbp]
\centering
\subfigure[]{
\begin{minipage}[t]{0.39\linewidth}
\centering
\includegraphics[width=1\textwidth]{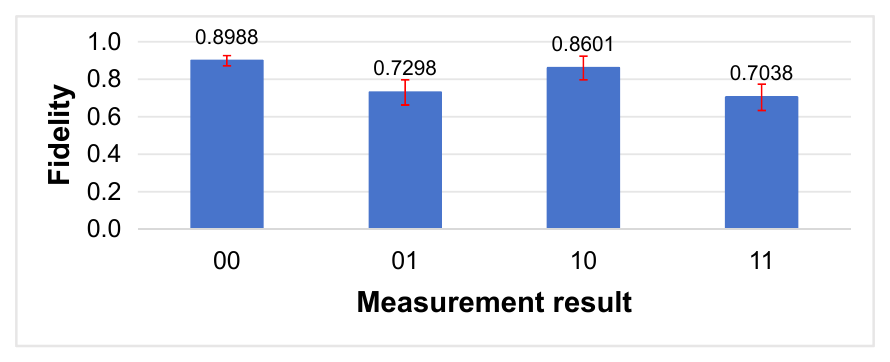}
\label{fig:17(b)}
\end{minipage}%
}%
\subfigure[]{
\begin{minipage}[t]{0.39\linewidth}
\centering
\includegraphics[width=1\textwidth]{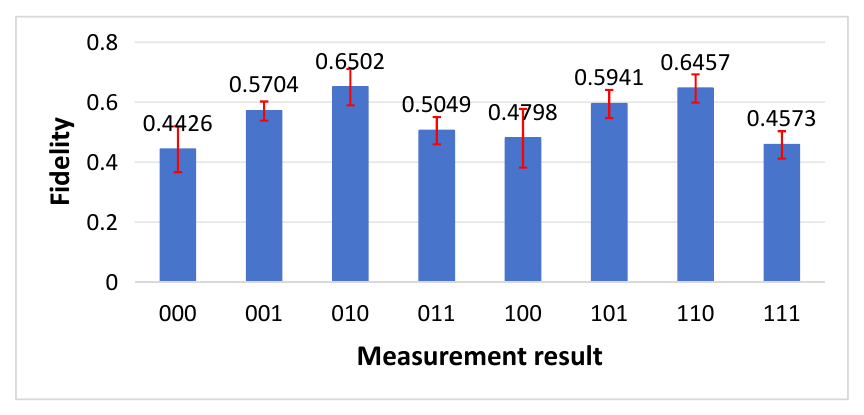}
\label{fig:18(b)}
\end{minipage}%
}%

\subfigure[]{
\begin{minipage}[t]{0.39\linewidth}
\centering
\includegraphics[width=1\textwidth]{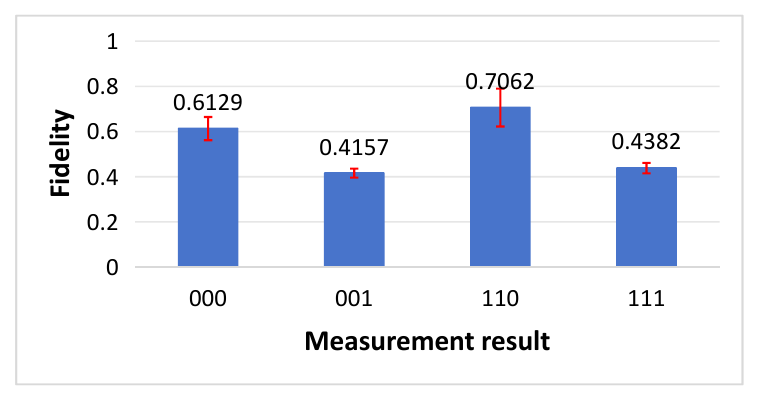}
\label{fig:19(b)}
\end{minipage}%
}%
\subfigure[]{
\begin{minipage}[t]{0.39\linewidth}
\centering
\includegraphics[width=1\textwidth]{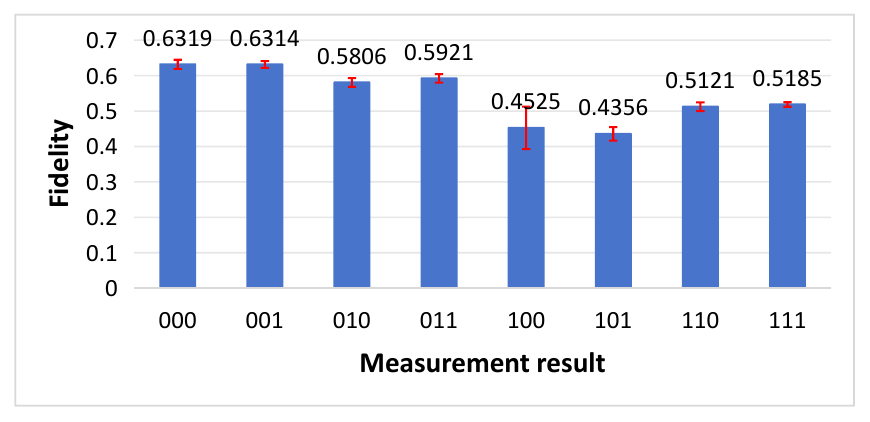}
\label{fig:20(b)}
\end{minipage}%
}%

\subfigure[]{
\begin{minipage}[t]{0.78\linewidth}
\centering
\includegraphics[width=1\textwidth]{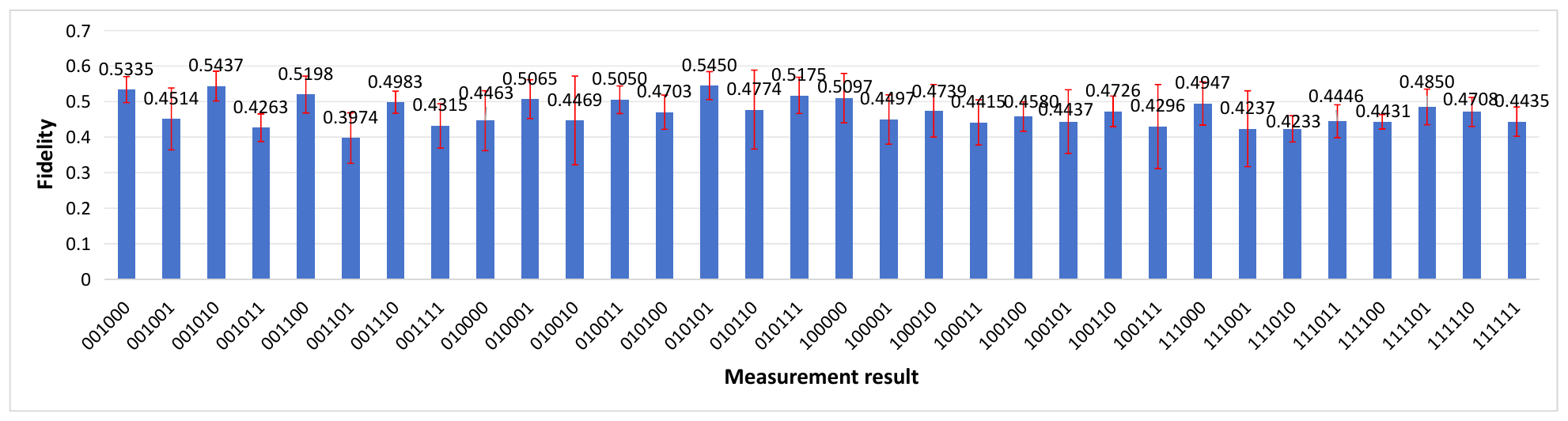}
\label{fig:22}
\end{minipage}%
}%
\centering
\caption{(a) The fidelity of the experimental quantum states of $q[0]$ and $q[3]$ corresponding to different measurements of $q[1]$ and $q[2]$ in the \Cref{fig:17(a)}. (b) The fidelity of the experimental quantum states of $q[0], q[4], q[5]$ corresponding to different measurements of $q[1], q[2]$ and $q[3]$ in in the \Cref{fig:18(a)}. (c) The fidelity of the experimental quantum states of $q[0], q[4]$ and $q[5]$ corresponding to different measurements of $q[1], q[2]$ and $q[3]$ in in the \Cref{fig:19(a)}. (d) The fidelity of the experimental quantum states of $q[0], q[3], q[5]$ corresponding to different measurements of $q[1], q[2]$ and $q[4]$ in in the \Cref{fig:20(a)}. (e) The fidelity of the experimental quantum states of $q[0], q[4], q[8]$ corresponding to different measurements of $q[1], q[2]$, $q[3], q[5], q[6]$ and $q[7]$ in in the \Cref{fig:21(a)}. Error bars are 1 SD calculated from all fidelity of 5 repetitive experimental runs.}
\end{figure*}

According to the module in \Cref{fig:9}, we can construct a quantum circuit of a $2$-dimensional GHZ state entanglement distribution using three GHZ states and recompile it into pulse form as shown in the \Cref{fig:21(a)} and \Cref{fig:21(b)} where $q[0], q[1], q[2], q[3], q[4], q[5], q[6], q[7], q[8]$ in quantum circuits are mapped to qubits $Q_1, Q_2, Q_3, Q_4, Q_6, Q_5, Q_7, Q_8$ on the real quantum processor. In the preparation stage, we prepared three GHZ states between $q[0], q[1], q[3]$ and $q[2], q[4], q[6]$ and $q[5], q[7], q[8]$ respectively. In the quantum walk stage, we performed the corresponding quantum walk between $q[1], q[2], q[3], q[5], q[6]$ and $q[7]$ according to the above theory. In the measurement stage, we measured $q[1], q[2], q[3], q[5], q[6]$ and $q[7]$ in proper basis and distribute a GHZ state between $q[0], q[4]$ and $q[8]$. We ran the quantum circuit with 10000 shots on superconducting quantum processor. Through the quantum state tomography, we can obtain the fidelity of the entangled state among $q[0], q[4]$ and $q[8]$ corresponding to any measurement result of $q[1], q[2]$, $q[3], q[5], q[6]$, $q[7]$. The fidelity results are shown in the \Cref{fig:22}. Through these fidelity, we can get average fidelity $F=47.0(0.7)\%$

\section{Discussion}

In scalable quantum networks, it is crucial to distribute multi-particle entangled states among the selected nodes, especially for high-dimensional entangled states. We developed a scheme for $d$-dimensional entanglement distribution based on quantum walks with coins. We present our $d$-dimensional entanglement distribution scheme for arbitrary quantum networks according to the entanglement distribution basic modules. We also experimentally implemented the basic modules of two-party and three-party entanglement distributions using Bell states or GHZ states on a superconducting quantum processor, and obtained high fidelities through quantum state tomography. As an application, we provide an example of constructing a quantum fractal network based on $d$-dimensional GHZ states and analyze some characteristics of its network structure, such as quantum transport properties and complex network properties.
The conditional shift operator in our quantum walk can be realized as a generalized CNOT quantum gate in quantum optical systems \cite{79}, which indicates that our framework for the entanglement distribution of high-dimensional quantum states is also highly feasible in experiments. In this study, to demonstrate the computational and experimental process, we employ quantum gate notation to characterize the quantum walk. However, this does not imply that the quantum walks in other physical systems must be experimentally implemented by realizing basic quantum gates. Quantum walks are often holistically implemented through specific physical processes. Therefore, our framework using quantum walks can be adapted to a broader range of physical experimental platforms.
Our entanglement distribution scheme not only circumvents the challenging task of performing joint Bell state measurements and high-dimensional multi-particle GHZ state measurements via quantum walks but also offers a more suitable framework for quantum repeaters in quantum networks not like \cite{34}. This provides a novel and comparatively straightforward implementation method for constructing larger-scale quantum networks in the future. Compared with the traditional 2-dimensional Bell state-based quantum repeater framework and its experimental realization, our framework enables the utilization of high-dimensional Bell states and GHZ states with any number of particles as basic entangled states. This offers additional design possibilities for future large-scale quantum network architecture, as well as a stronger theoretical foundation for combating quantum noise and facilitating efficient communication. In the future, we will consider distributing other multi-particle entangled states in arbitrary quantum network nodes, such as Dicke states.

\section*{Data availability statement}
The data that support the findings of this study are available on request from the corresponding author upon reasonable request.

\section*{Acknowledgments}
The research is supported by the National Key R\&D Program of China under Grant No. 2023YFA1009403, the National Natural Science Foundation special project of China (Grant No.12341103), the National Natural Science Foundation of China (Grant No.62372444), the National Natural Science Foundation of China (Grant No.92265207), the National Natural Science Foundation of China (Grant No.T2121001).

\section*{References}

\bibliographystyle{unsrt} 
\bibliography{ref}

\section*{Appendix}
\appendix

\section{Properties of the fractal network}

When we construct a quantum fractal network based on the Sierpinski gasket, a tripartite channel is established between the three nodes where particles of the same GHZ state are located. The method of constructing channels based on entangled states determines the properties of the fractal networks that we constructed. We analyzed these properties from the perspective of complex networks. 

Because our fractal network is designed based on the Sierpinski gasket structure, its fractal dimension is $\frac{\ln3}{\ln2}$. 
After the calculation, the number of vertices $N(t)$ in $F(t)$ is $\frac{3^{t+1}+3}{2}$, and the number of edges $E(t)= \frac{3^{t+2}-3}{2}$. Therefore, for a large $t$, the average degree $\overline{k(t)}= \frac{2E(t)}{N(t)}$ is approximately six, indicating that this is a sparse network\cite{51} similar to some actual networks. 

At $F(t_i), t_i>0$ , the new vertex $i$ in $F(t)$ has a degree of $k=4(t-t_i+1)$. The three nodes in $F(0)$ do not affect the degree distribution of the network when $t$ is large; therefore, they can be disregarded. Therefore, the cumulative degree distribution $P(k)= \frac {3 ^ {t_i+1}-3} {3^{t+1}+3}$. Set $t_i=t-\frac {k}{4}+1 $ and $t$ tends to infinity, and $P(k)=3\cdot e^{-\frac {\ln{3}\cdot k}{4}}$. From this, we can see that the degree distribution of this network meets the exponential distribution, not the power law distribution.

For vertex $i$ with $k$ in the middle of $F(t)$ above, the number of edges between adjacent points  $e=4+\frac{3}{2}(k-4)$, and $k=4(t-t_i+1)$, the average clustering coefficient $C(t)$ of $F(t)$ is
\begin{eqnarray}
\sum\limits_{t_i=1}^{t}\frac{3^{t_i}}{3^{t+1}+3}\cdot\frac{4e}{k(k-1)}+\frac{6}{3^{t+1}+3}\cdot\frac{3t+1}{(t+1)(2t+1)}.
\end{eqnarray}
When $t$ tends to infinity, $C(t)$ tends to a non-zero constant of 0.5480. Therefore, we know that the network has a small world property\cite{56} through a larger clustering coefficient and a linear relationship between the average path length and the logarithm of the number of vertices, which generally exist in actual complex networks.

\section{Experimental setup}

\begin{figure*}[ht]
\centering
\includegraphics[width=0.85\textwidth]{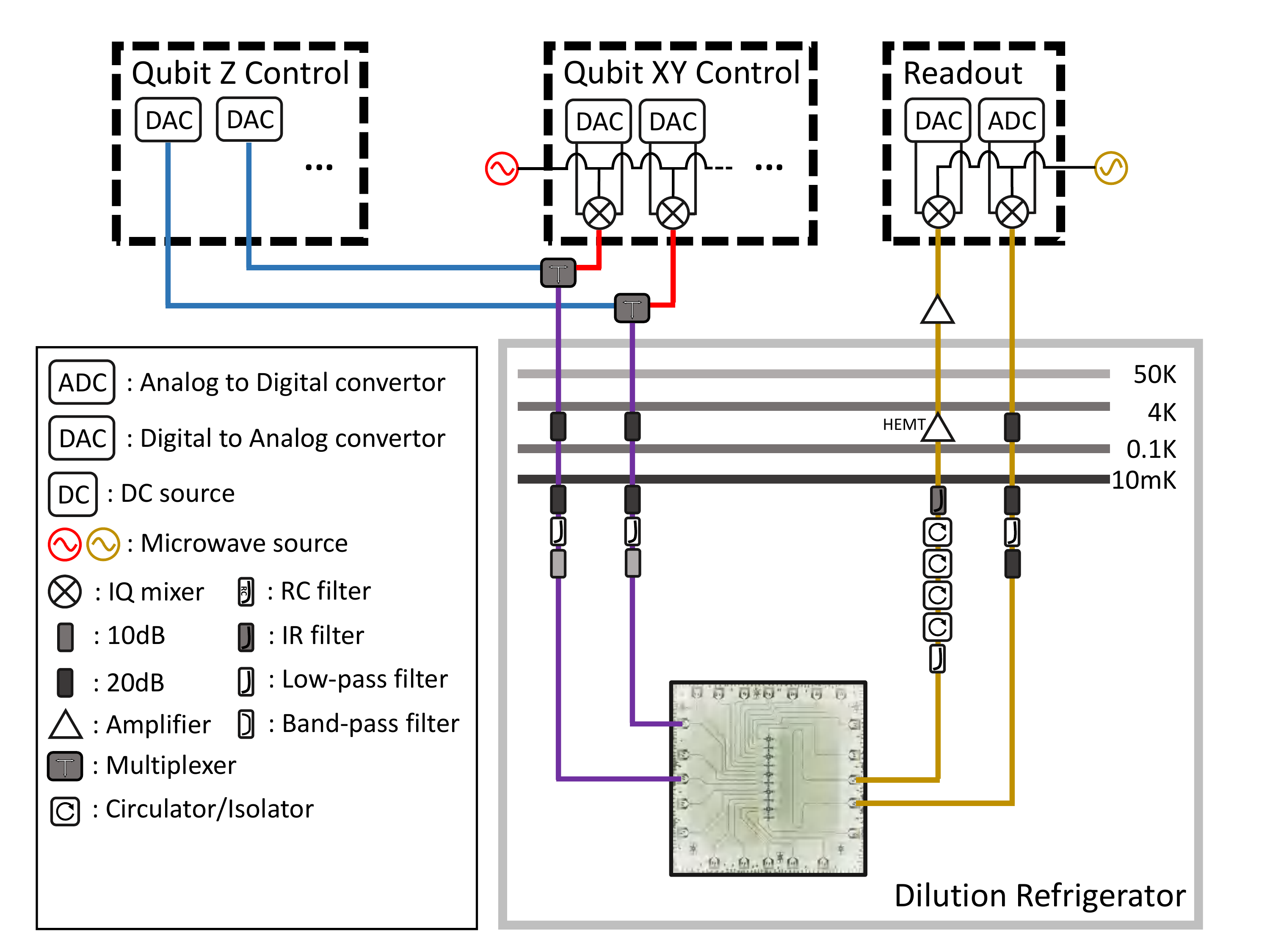}
\caption{Wiring of cryogenic and control electronics equipment. The QPU is
installed at 10 mk stage of the dilution refrigerator. We merge fast Z control line and XY control line of each qubit into one outside the dilution refrigerator. For readout, all qubits can be measured simultaneously.}
\label{wiring}
\end{figure*}

In this experiment, the quantum processor consisted of 10 superconducting qubits, with neighboring qubits coupled through capacitance, and all the qubits were arranged in a one-dimensional chain. The quantum processor works in a low-temperature environment of approximately 10 mk formed by cryogenic equipment, and each qubit requires one microwave channel to control its frequency, and another microwave channel is needed for excitation. A group of control electronics working at room-temperature, such as arbitrary waveform generators and microwave sources, provides microwave pulses, which are connected to quantum chips through microwave transmission lines. In addition, a microwave channel is required to transmit readout pulses, and a microwave transmission line is used to simultaneously measure the states of all qubits. The scattered signal was demodulated by the ADC to obtain the measurement results. The wiring of the control electronics and cryogenic equipment is illustrated in the \Cref{wiring}.

Owing to fixed coupling rates, the difference in idle frequencies between adjacent qubits should be large enough to minimize residual Z-Z coupling between qubits with large detuning, also known as the AC Stark effect. When applying readout pulses to each qubit simultaneously, each qubit was biased at a designed point to reduce the correlated readout errors caused by the measurement.

Some parameters of the quantum processor are listed in \Cref{para} and \Cref{CZ}. The fidelities of the two-qubit CZ gates were obtained using quantum process tomography.The qpt fidelity is obtained when the other qubits are in the ground state.

\begin{table*}[ht]
    \centering
    \setlength{\tabcolsep}{1.5mm}{
    \begin{tabular}{ccccccccccc} 
        \hline Qubit&$Q_{1}$&$Q_{2}$&$Q_{3}$&$Q_{4}$&$Q_{5}$&$Q_{6}$&$Q_{7}$&$Q_{8}$&$Q_{9}$&$Q_{10}$\\
        \hline $\omega^{s}_{j}/2\pi$ (GHz)  &5.536&5.069 &5.660 &4.742 &5.528 &4.929 &5.451 &4.920 &5.540 &4.960 \\
        $\omega_{10,j}/2\pi$  (GHz)    &5.310 &4.681 &5.367 &4.702 &5.299 &4.531 &5.255 &4.627 &5.275 &4.687 \\
        $\omega^{r}_{j}/2\pi$  (GHz) & 5.493 &4.800 &5.455 &4.734 &5.300 &4.436 &5.252 &4.807 &5.309 &4.402 \\
        $\eta_{j}/2\pi$   (GHz) & 0.250 &0.207 &0.251 &0.206 &0.251 &0.203 &0.252 &0.204 &0.246 &0.208 \\
        $g_{j,j+1}/2\pi$ (MHz) & 12.07 &11.58 &10.92 &10.84 &11.56 &10.00 &11.74 &11.70 &11.69 & - \\
        $T_{1,j}$ (us)    &30.1 &17.6 &23.2 &10.7 &33.3 &61.0 &31.0 &33.7 &25.6 &75.2 \\
        $T_{2,j}^{*}$ (us)    &2.14 &1.17 &1.56 &2.67 &1.89 &1.30 &2.07 &1.50 &1.56 &2.30 \\
        $F_{0,j}$ (\%) & 98.50 &98.45 &97.40 &98.13 &97.47 &96.43 &96.60 &96.37 &98.17 &98.00 \\
        $F_{1,j}$ (\%) & 94.20 &93.97 &94.80 &94.27 &92.03 &86.80 &93.03 &90.97 &92.17 &88.30 \\
        \hline
    \end{tabular}}
    \caption{Device parameters. $\omega^{s}_{j}$ represents the maximum frequency of $Q_{j}$. $\omega_{10,j}$ corresponds to the idle frequency of $Q_{j}$. $\omega^{r}_{j}$ indicates the resonant frequency of $Q_{j}$ during readout. $\eta_j$ shows the anharmonicity of $Q_{j}$. $g_{j,j+1}$ is the coupling strength between nearest-neighbor qubits. $T_{1,j}$ and $T^{*}_{2,j}$ represent the relaxation time and coherence time of $Q_{j}$. $F_{0,j}$ and $F_{1,j}$ are readout fidelities of $Q_{j}$ in $|0\rangle$ and $|1\rangle$, which can be used for readout correction.}
    \label{para}
\end{table*}

\begin{table*}[ht]
    \centering
    \setlength{\tabcolsep}{0.3mm}{
    \begin{tabular}{cccccccccc} 
        \hline  & $Q_{1}Q_{2}$ & $Q_{2}Q_{3}$ & $Q_{3}Q_{4}$ & $Q_{4}Q_{5}$ & $Q_{5}Q_{6}$ & $Q_{6}Q_{7}$ & $Q_{7}Q_{8}$ & $Q_{8}Q_{9}$ & $Q_{9}Q_{10}$ \\
        \hline $T_{\mathrm{CZ}} (\mathrm{ns})$ & 48.9 & 46.1 & 40.1 & 39.0 & 42.8 & 41.8 & 43.6 & 40.5 & 41.1 \\
        $P_{\mathrm{ref}}$ (\%) & 97.98 & 98.45 & 95.74 & 96.64 & 98.29 & 98.24 & 97.06 & 96.13 & 96.19 \\
        $P_{\mathrm{target}}$ (\%) &96.02 &96.16 &95.24 &95.30 &94.63 &95.50 &94.04 &92.11 &94.42 \\        
        $F_{RB}$ (\%) &98.50(1) &98.25(1) &99.61(1) &98.95(1) &97.21(1) &97.91(2) &97.67(1) &96.87(1) &98.62(1) \\
        \hline 
    \end{tabular}}
    \caption{Gate fidelities of CZ gates. $T_{\mathrm{CZ}}$ represents the total duration of the pulse forming CZ gate. $P_{\mathrm{ref}} (P_\mathrm{target})$ represents depolarization rate obtained from randomly sampled reference(target) sequences. $F_{RB}$ is the process fidelity of CZ gate obtained from randomized benchmarking.}
    \label{CZ}
\end{table*}

\section{Crosstalk correction}

In this experiment, the Z control signal and XY control signal of each quantum bit on the quantum processor were combined into a single signal at room temperature and transmitted to the quantum bit through a transmission line. There is classical electromagnetic crosstalk between the control signals corresponding to each quantum bit, and the magnitude of the Z signal crosstalk increases nonlinearly with the increase in signal application time. The characteristic time is about 10us, This low-frequency crosstalk is caused by inductive coupling between the loops composed of Z-signals from different channels, which can lead to a temporal correlation in the fidelity of the two bit quantum logic gates used in the experiment. Currently, this crosstalk can be greatly reduced through the use of flip soldering packaging technology. However, in this processor, we need to address the issue of crosstalk from the perspective of measurement and control software. As crosstalk changes over time, we cannot perform linear correction through the traditional use of crosstalk matrices. Instead, we measure the response function of bits to the Z-signal input of all other channels, apply compensation pulses, and mitigate Z-signal crosstalk between channels.

In addition to the classic electromagnetic crosstalk of low-frequency signals, such as the Z signal, crosstalk of high-frequency signals, such as the XY signal in quantum chips, also exists. In the experiment, we used the following method to determine the degree of XY crosstalk. Taking $Q_1$ and $Q_2$ as an example, in the first experiment, we did not operate on $Q_1$ and conducted Ramsey testing on $Q_2$. In the second experiment, as a control, we excite $Q_1$ from $|0\rangle$ to $|1\rangle$ through XY signal, and then conduct Ramsey testing on $Q_2$. When the periods of the two Ramsey curves are equal but the phases are not, this indicates the existence of classical electromagnetic crosstalk in the XY signal, and the magnitude of the phase difference indicates the strength of crosstalk. In this case, we increased the length of the XY pulse for pulsing $Q_1$ and decreased the pulse amplitude to reduce classical XY crosstalk.

\section{Readout correction}

Because of the interaction between microwave pulses used for dispersion reading and quantum bits, the scattered reading pulses detected by the DAC often cannot accurately reflect the state of the quantum bits. However, we can correct this part of the errors introduced by reading through post-processing. First, we ensured that there was no correlation between the read errors by changing the frequency of the qubits when applying a read pulse. Then, we measure the probability of each quantum bit being in the $|0\rangle$ state, and write it as $F_0^i$; Then we measure the probability that the result obtained when each quantum bit is in a $|1\rangle$ state is written as $F_1^i$, which is called read fidelity.

After the above operations, we can obtain the measurement transfer matrix $M_i$ for the i-th quantum bit $Q_i$, 
\begin{eqnarray}
M_{j}=
\left(
\begin{matrix}
F^{j}_{0}& 1-F^{j}_{1}\\
1-F^{j}_{1}& F^{j}_{1}   
\end{matrix}
\right)
\end{eqnarray}
which represents the mapping relationship between the ideal measurement result of a single quantum bit and actual measurement result.

The measurement transfer matrix for multiple quantum bits can be written in the following form:
\begin{eqnarray}
M = M_{0} \otimes M_{1} \otimes \cdots \otimes M_{N},
\end{eqnarray}

Therefore, we can eliminate errors introduced by reading by inverting the measurement transfer matrix of multiple qubits. If the measured probability of the qubit string before correction is $\tilde{\textbf{P}}$, the ideal probability vector obtained from the measurement is
\begin{eqnarray}
{\textbf{P}}=M^{-1} \tilde{\textbf{P}}.
\end{eqnarray}

\end{document}